\pdfoutput=1
\documentclass{jfm} %[draft] or [final]
\usepackage[utf8]{inputenc}

\usepackage{lmodern}
\usepackage[T1]{fontenc}

\usepackage{mystyle}

\title[Scattering of internal tides]{Scattering of internal tides by barotropic quasigeostrophic flows}
\author[M. A. C. Savva \& J. Vanneste]{Miles A. C. Savva and Jacques Vanneste} 
\affiliation{School of Mathematics and Maxwell Institute for Mathematical Sciences, \\
University of Edinburgh, Edinburgh EH9 3FD, UK}

\begin{document}
\maketitle
\begin{abstract}
Oceanic internal tides and other inertia-gravity waves propagate in an energetic turbulent flow whose lengthscales are similar to the  wavelengths.
Advection and refraction by this flow cause the scattering of the waves, redistributing their energy in wavevector space. As a result, initially plane waves radiated from a source such as a topographic ridge become spatially incoherent away from the source. To examine this process, we derive a kinetic equation which describes the statistics of the scattering under the assumptions that the flow is quasigeostrophic, barotropic, and well represented by a stationary homogeneous random field. Energy transfers are quantified by computing a scattering cross section and shown to be restricted to waves with the same frequency and identical vertical structure, hence the same horizontal wavelength. For isotropic flows, scattering leads to an isotropic wavefield. We estimate the characteristic time and length scales of this isotropisation, and study their dependence on parameters including the energy spectrum of the flow. Simulations of internal tides generated by a planar wavemaker carried out for the linearised shallow-water model confirm the pertinence of these scales. A comparison with the numerical solution of the kinetic equation demonstrates the validity of the latter and illustrates how the interplay between wave scattering and transport shapes the wave statistics.

\end{abstract}
%\tableofcontents

\section{Introduction}

The propagation of inertia-gravity waves in the ocean has received a great deal of attention, mainly motivated by the role they play in the large- and mesoscale circulation, through wave--mean-flow interaction, mixing and dissipation. The inertia-gravity-wave spectrum is dominated by two types of waves: near-inertial oscillations, with frequencies close to the inertial frequency $f$, which are mainly generated  by winds, and internal tides (ITs), primarily at the semi-diurnal lunar frequency, which are generated by the interaction of the barotropic tide with topography \citep[e.g.,][]{ferr-wuns09}. 
Near-inertial oscillations have distinctive dynamics, including weak dispersion and weak vertical motion, that stem from their unique place at the low-frequency end of the inertia-gravity wave spectrum \citep{alfo-et-al}; ITs, in contrast, are generic mid-frequency inertia-gravity waves with their externally imposed frequency as their sole defining property. 

The ocean's highly energetic quasigeostrophic turbulence has a strong impact on the structure of both inertial oscillations and ITs and hence, in the case of ITs, on their signature on the sea-surface height \citep{rainvillepinkel,rayzaron}. There is by now an extensive literature devoted to this impact, with a recent impetus  provided by upcoming high-resolution satellite-altimetry instruments and the need to disentangle ITs from mesoscale (balanced) motion in the observed sea-surface height. We refer the reader to the recent papers by \cite{wagner_ferrando_young_2017} and \cite{dunphy} for further background.

A key aspect of the interactions between quasigeostrophic turbulence and both near-inertial oscillations and low-mode ITs is that turbulence and waves  share similar horizontal scales, of the order of 100 km.
A consequence is that, in such a regime, the WKB approximation on which much of our understanding of inertia-gravity-wave propagation is built is not valid. This has prompted the development of simplified, wave-averaged models that rely only on time-scale separation to represent the interactions between waves and flow in a simplified manner.  Models of this kind include the Young--Ben Jelloul model of near-inertial oscillations in a quasigeostrophic flow \citep{youn-benj} and its extensions accounting for the feedback of the waves on the flow \citep{xie,wagner_young_2016,thomas_smith_buhler_2017}. \citet{wagner_ferrando_young_2017} recently derived an analogue of the Young--Ben Jelloul model equation for ITs. This model is formulated in physical space and retains a stiff term which  enforces the constraint that the ITs' fixed frequency imposes on their spatial structure and which  cannot be eliminated without resorting to a Fourier-space formulation.

The present paper  focuses similarly on the impact of a quasigeostrophic flow on ITs, making no asymptotic assumption about the relative horizontal scales of ITs and flow. Our focus is on quantifying the scattering induced by  a barotropic (i.e.\ $z$-independent) geostrophic turbulent flow which we model as a spatially homogeneous random field. We take advantage of the  assumption of barotropic flow to use a vertical-mode expansion and thus reduce the problem to the study of an (equivalent) shallow-water model. 
By applying the theory of waves in random media developed by \cite{ryzhik}, we derive a kinetic equation describing, in a statistically averaged sense, the energy exchanges between ITs with different wavevectors. The theory is formulated in terms of a wavevector-resolving energy density, $a(\vec{x},\vec{k},t)$, which makes it possible to capture spatial variations of the wave energy. The form of the scattering term in the kinetic equation for  $a(\vec{x},\vec{k},t)$ shows that energy transfers are restricted to waves with the same frequency or, equivalently, the same wavenumber  $|\vec{k}|$. These transfers result from  interactions within resonant triads consisting of two ITs of equal frequencies with a zero-frequency flow (vortical) mode -- the so-called catalytic interactions of \citet{lelong} and \cite{bartello}.
The rate of these transfers is proportional to the energy spectrum of the geostrophic flow. In the case of an isotropic flow, the scattering leads to the relaxation of the energy density towards a locally isotropic density $a(\vec{x},|\vec{k}|)$. Our theory complements that developed by \cite{ward},  shifting from a deterministic to a statistical treatment that can be regarded  as a version of wave turbulence \citep{naza} in which the statistics of the flow are prescribed. 
 It  generalises the theory developed by \cite{danioux} for near-inertial oscillations to inertia-gravity waves of arbitrary frequencies. Note that our statistical approach focuses on a homogeneous field of scatterers (resulting from the turbulent flow) and that different analysis techniques apply to waves incident on isolated scatterers \citep[e.g.][]{olbers81}. 

We analyse the predictions of the kinetic equation, focusing our attention on parameters representative of the first baroclinic mode of the semidiurnal lunar tide $M_2$. These predictions include a time scale for wave isotropisation applicable to statistically homogeneous wavefields (i.e.\ such that $\nabla_{\vec{x}} a =0)$ and in particular to the isotropisation of an initially plane wave examined numerically by \citet{ward} in a shallow-water setup.
The kinetic equation applies to more general, non-homogeneous situations in which the energy density is modulated spatially ($\nabla_{\vec{x}} a \not=0$). This makes it possible to study the scattering of ITs generated by a localised source such as a topographic ridge. \cite{ponte} and \cite{dunphy} recently used three-dimensional Boussinesq simulations to study this problem and quantify the temporal incoherence of the ITs that results from the presence of a time-dependent turbulent flow.  (See also \citet{kelly1} and \citet{kelly2} for simulations of ITs in realistic configurations.) We carry out shallow-water simulations in a setup analogous to theirs, and compare the results with direct solutions of the kinetic equation. This provides an estimate for the length scale over which the wave field becomes isotropic and, more broadly, sheds light on the interplay between transport of the wave energy by the group velocity and scattering.
We emphasise that our theory concentrates on the statistical properties of the IT energy and makes no predictions for their phase. In the regime we consider, with a flow assumed to vary on a timescale much larger than the tidal period, the stationarity of the turbulent energy spectrum ensures that the tidal energy remains concentrated at the single wavenumber dictated by the fixed tide frequency.

The plan of the paper is as follows. We describe the equations satisfied by linear internal waves propagating on a barotropic quasigeostrophic flow in \S\ref{sec:scattering}, expanding them in vertical modes to obtain an equivalent shallow-water system for each mode. We then sketch the derivation of the kinetic equation using the method of \cite{ryzhik}, relegating the technical computations to Appendix \ref{app:kinetic}. We focus on the application of the kinetic equation to the case of an isotropic flow in \S\ref{sec:isotropic} where we derive explicit estimates for the time- and lengthscales over which the IT field becomes isotropic. In \S\ref{sec:simulations} we compare our theoretical predictions with direct simulations of the linearised shallow-water equations and with numerical solutions of the kinetic equation itself in a configuration where a wavemaker forces a plane IT in a turbulent flow. We conclude in \S\ref{sec:discussion} with a discussion.

\section{Scattering theory for internal tides} \label{sec:scattering}

\subsection{Model}

We model the propagation of ITs through a turbulent quasigeostrophic eddy field using the hydrostatic Boussinesq equations linearised about a slowly evolving barotropic flow.
The background flow is time dependent and geostrophically and hydrostatically balanced, given by $\vec{U}=(U,V,0)=(-\partial_y\psi,\partial_x\psi,0)$ in terms of a $z$-independent streamfunction $\psi$. With these assumptions, the  linearised hydrostatic--Boussinesq equations read
\begin{equation}
\begin{aligned}
    \partial_t\vec{u}+\vec{U}\cdot\nabla\vec{u}+\vec{u}\cdot\nabla\vec{U}+f\hat{\vec{z}}\times\vec{u}&=-\nabla p,\\
    \partial_z p&=b,\\
    \partial_t b+\vec{U}\cdot\nabla b+N^2w&=0,\\
    \nabla\cdot\vec{u}+\partial_zw&=0,
\end{aligned}\label{boussinesq1}
\end{equation}
where $(\vec{u},w)$ denotes the IT velocity, $\hat{\vec{z}}$ is the vertical unit vector,
$p$ is the pressure normalised by a reference density, $b$ the buoyancy, $f$ the Coriolis parameter, and $N(z)$ the buoyancy frequency. We use the notation $\nabla=(\partial_x,\partial_y,0)$ for the horizontal gradient throughout. 

Assuming a flat bottom boundary and rigid lid, we project \eqref{boussinesq1} onto baroclinic modes to obtain a set of rotating shallow-water equations governing their amplitudes:
\begin{equation}
    \begin{aligned}
    \partial_t\vec{u}_m+\vec{u}_m\cdot\nabla\vec{U}+\vec{U}\cdot\nabla\vec{u}_m+f\hat{\vec{z}}\times\vec{u}_m&=-g\nabla\eta_m,\\
    \partial_t\eta_m+\vec{U}\cdot\nabla\eta_m+h_m\nabla\cdot\vec{u}_m&=0,
    \end{aligned}\label{SWEs}
\end{equation}
where $\eta_m$ is the equivalent surface height and $h_m$ is the equivalent depth (see Appendix \ref{app:vmode-decomp} for details). Note that this system differs from the one obtained by linearising the shallow-water equations about a background flow in geostrophic balance since the latter system includes a contribution from the (sloping) background free surface. 

For physical applications in later sections, we take parameters corresponding to the first baroclinic mode only, since this contains the majority of the IT energy.
In the ocean, energy is transferred between vertical modes as a result of vertical shear. However, as discussed by \citet{dunphy-lamb} and \citet{ponte} the effect is small. Simulations in \cite{dunphy} put the transfer of energy from the first mode to higher modes at 3\% in their most extreme cases, with a highly energetic background flow, and less for typical ocean conditions. We drop the subscript $m$ in \eqref{SWEs} from this point on.

\subsection{Derivation of the kinetic equation} 

We study IT scattering in the distinguished regime where the spatial scale of the flow, $L_*$ say, is of the same order as the wavelength, that is, $\abs{\vec{k}}L_* = O(1)$, where $\vec{k}=(k,l)$ is the IT horizontal wavevector. The assumption of a geostrophic flow requires a small Rossby number $\Ro=U_*/(fL_*)\ll 1$, where $U_*$ is a typical flow velocity; in turn, this implies that the background flow velocities  are small compared with the wave phase speed $\omega/|\vec{k}|$, where
\begin{equation}
\omega = \sqrt{f^2 + gh| \vec{k}|^2}
\label{disp_relation}
\end{equation}
is the IT frequency, since $U_*/(\omega/|\vec{k}|)=O(U_* /(\omega L_*))=O(\Ro)$, given that  $\omega=O(f)$ away from the equator. With the flow timescale $T_*$ taken as the natural advective timescale $L_*/U_*$, this also implies that the flow evolves slowly compared with the IT timescale since $\omega T_*=O(\Ro) \ll 1$. We further assume that, while the IT phases vary over the lengthscale $|\vec{k}|^{-1}$, their amplitudes vary over a much larger scale $\varepsilon^{-1} \abs{\vec{k}}$, where $\varepsilon \ll 1$. We adopt the scaling $\varepsilon=O(\Ro^2)$. As emerges below, this is the distinguished scaling that ensures that transport and scattering affect the wave field at the same order and are both captured at leading order by our asymptotic model.

Since our focus is on generic, statistical properties of the IT field, we model the turbulent background flow by a random streamfunction with homogeneous and stationary statistics. With our scaling assumptions, it is then possible to derive a single equation that describes the scattering and transport of IT energy following the theory of \cite{ryzhik}. 
This theory is formulated in terms of the Wigner transform
\begin{equation}
    W(\vec{x},\vec{k},t)=\frac{1}{(2\pi)^2}\int_{\mathbb{R}^2}\exp^{\i \vec{k}\cdot\vec{y}}\vec{\phi}(\vec{x}-{\vec{y}}/{2},t)\vec{\phi}^*(\vec{x}+{\vec{y}}/{2},t)\d\vec{y},\label{Wigner_definition1}
\end{equation}
where $\vec{\phi}=(u,v,\eta)^\mathrm{T}$ groups the dynamical fields and the asterisk denotes conjugate transpose. The Wigner transform is a Hermitian matrix; it is not necessarily positive definite, although its integral over the wavevector $\vec{k}$, simply given by $\vec{\phi}(\vec{x},t) \vec{\phi}^*(\vec{x},t)$, is.  

The equation derived by Ryzhik et al.\ (1996) governs the evolution of a scalar amplitude $a(\vec{x},\vec{k},t)$ which appears naturally in an eigenvector decomposition of the matrix $W(\vec{x},\vec{k},t)$ (see Appendix \ref{app:kinetic} for details). Physically, this amplitude is interpreted as a wavevector-resolving energy density, related to the (leading-order) energy density of the system by
\begin{equation}
    \mathcal{E}_0(\vec{x},t)=\frac{1}{2}\int_{\mathbb{R}^2}a(\vec{x},\vec{k},t)\d\vec{k}.\label{en_density}
\end{equation}
To avoid any confusion, we emphasise that $a(\vec{x},\vec{k},t)$ itself represents a wave-energy density and not a wave amplitude; as its definition in \eqref{W0} makes clear, it is a quadratic function of the wave fields, like the Wigner transform $W(\vec{x},\vec{k},t)$.

In Appendix \ref{app:kinetic}, we show that $a(\vec{x},\vec{k},t)$ satisfies the kinetic equation
\begin{equation}
    \partial_t a+\nabla_\vec{k}\omega\cdot\nabla_\vec{x}a=\mathcal{L}a-\Sigma a,\label{transport_equation}
\end{equation}
where the notation $\nabla_{\vec{x}}=\nabla=(\partial_x,\partial_y)$ emphasises that the spatial gradient applies to functions of both $\vec{x}$ and $\vec{k}$, and where $\nabla_{\vec{k}}=(\partial_{k},\partial_{l})$.
Here $\omega$ is determined by the IT dispersion relation \eqref{disp_relation}, so that $\nabla_{\vec{k}} \omega$ is the group velocity and the left-hand side of \eqref{transport_equation} represents the familiar wave transport. (The term $-\nabla_{\vec{x}} \omega \cdot \nabla_{\vec{k}} a$ would be added if $\omega$ depended explicitly on $\vec{x}$.) The right-hand side represents wave scattering by the background flow. The first term, given by
\begin{equation}
    \mathcal{L}a(\vec{x},\vec{k},t)=\int_{\mathbb{R}^2}\sigma(\vec{k},\vec{k}')a(\vec{x},\vec{k}',t)\d\vec{k}',\label{La}
\end{equation}
quantifies the transfers of energy from all wavevectors $\vec{k}'$ into wavevector $\vec{k}$ that result from interactions with the background flow; the second term, where
\begin{equation}
    \Sigma=\Sigma(\vec{k})=\int_{\mathbb{R}^2}\sigma(\vec{k},\vec{k}')\d\vec{k}',\label{Sigma}
\end{equation}
is the total scattering cross section,
quantifies the energy lost by  wavevector $\vec{k}$ to all other wavevectors.

The function $\sigma(\vec{k},\vec{k}')$ that appears in \eqref{La}--\eqref{Sigma} is the main object of interest. It is known as the differential scattering cross section and measures the rate at which energy is scattered from $\vec{k}$ to $\vec{k}'$ at a position $\vec{x}$ in space. 
%{Both $\sigma$ and $\Sigma$ are nonnegative and $\sigma$ is usually symmetric in $\vec{k}$ and $\vec{k}'$.} For the shallow water system \eqref{SWEs} we have the explicit formula for $\sigma$ given by
We obtain it in the form
\begin{multline}
    \sigma(\vec{k},\vec{k}')=\frac{2\pi}{gh{\omega^3|\vec{k}|^5}}\Big\{\abs{\vec{k}\times\vec{k}'}^2\big[(\omega^2+f^2)\vec{k}\cdot\vec{k}'-f^2\abs{\vec{k}}^2\big]^2\\
    +f^2\omega^2\big[\abs{\vec{k}\times\vec{k}'}^2+\vec{k}\cdot\vec{k}'(\abs{\vec{k}}^2-\vec{k}\cdot\vec{k}')\big]^2\Big\}\frac{\hatt{E}(\vec{k}-\vec{k}')}{\abs{\vec{k}-\vec{k}'}^2}\delta(\abs{\vec{k}}-\abs{\vec{k}'}),\label{cross_section}
\end{multline}
where $\vec{k}\times\vec{k}'=k l'-l k'$, and $\hatt{E}$ is the energy spectrum of the flow. We note that $\sigma(\vec{k},\vec{k}')$ is real, positive, and symmetric with respect to the exchange between $\vec{k}$ and $\vec{k}'$. In Appendix \ref{app:kinetic}, we also show that these properties ensure conservation of the leading-order energy density \eqref{en_density}: 
\begin{equation}
    \partial_t\mathcal{E}_0+\nabla_\vec{x}\cdot\mathcal{F}_0=0,
\end{equation}
where 
\begin{equation}
    \mathcal{F}_0(\vec{x},t)=\int_{\mathbb{R}^2}\nabla_\vec{k} \omega(\vec{k}) \, a(\vec{x},\vec{k},t)\d\vec{k}
\end{equation}
is the leading-order energy flux (see \eqref{energy_cons}). 

The presence of the factor $\delta(\abs{\vec{k}}-\abs{\vec{k}'})$ in \eqref{cross_section} implies that energy is only exchanged between wavevectors of the same magnitude, that is, between waves with the same frequency, as a result of the assumed slow time dependence of the background flow.  Thus, in the regime considered, the IT energy is confined to the constant-frequency circle $\abs{\vec{k}}=((\omega^2-f^2)/(gh))^{1/2}$ in the wavevector plane. 
This can be related to the observation that the background flow only enters $\sigma(\vec{k},\vec{k}')$ through its energy spectrum $\hatt{E}$, which, for the statistically stationary flows considered, is time independent. The scattering described by \eqref{transport_equation} results from the resonant interactions of two ITs, with wavevectors $\vec{k}$ and $\vec{k}'$ and identical frequencies, with a vortical flow mode of wavevector $\vec{k}-\vec{k'}$ and zero frequency. Because of potential-vorticity conservation, these interactions would leave the vortical mode unaffected even if it were allowed to evolve freely; hence they have been termed catalytic interactions \citep{lelong,bartello,ward}. We emphasise that \eqref{transport_equation} captures the net effect of multiple triadic interactions acting over long time scales. This is why the time scale of evolution is not  linear in the flow amplitude but quadratic, dictated by the energy spectrum of the flow, in a manner familiar from wave turbulence \citep[e.g.][]{naza}.

\section{Scattering in isotropic turbulence} \label{sec:isotropic}

\subsection{Isotropisation}

In this section we use the kinetic equation   \eqref{transport_equation} to make  predictions about the scattering process and quantify the time and length scales over which ITs lose their spatial coherence. For simplicity, we assume that the flow is isotropic, $\hatt{E}(\vec{k})=\hatt{E}(\abs{\vec{k}})$. This motivates the use of polar coordinates for the wavevector, such that 
\begin{equation}
    \vec{k}=\abs{\vec{k}}\left( \begin{array}{c}
        \cos \theta     \\
         \sin \theta
    \end{array} \right) \ \ \textrm{and} \ \ \vec{k}'=\abs{\vec{k'}}\left(\begin{array}{c}
        \cos(\theta+\theta ')  \\
         \sin(\theta + \theta ')
    \end{array}\right),
\end{equation}
{where $\theta'$ is the angle between $\vec{k}$ and $\vec{k}'$. }
The change of coordinates reduces the scattering operator \eqref{La} to
\begin{equation}
    \mathcal{L}a(\vec{x},\abs{\vec{k}},\theta,t)=\int_{-\pi}^\pi\sigma'(\abs{\vec{k}},\theta')a(\vec{x},\abs{\vec{k}},\theta-\theta',t)\d\theta',\label{La_polar}
\end{equation}
where 
\[
\sigma'(\abs{\vec{k}},\theta'):=\int_0^\infty \sigma(\vec{k},\vec{k}') \, \abs{\vec{k}'}\, \d\abs{\vec{k}'}.
\]
Note that we have used the evenness of $\sigma'$ in $\theta'$ to rewrite \eqref{La_polar} as a convolution. Note also that $\sigma'$ is independent of the direction $\theta$ of $\vec{k}$ because the scattering process is rotationally invariant. The scattering cross section \eqref{cross_section}
can be written explicitly as
\begin{equation}
\sigma'(\abs{\vec{k}},\theta)=\frac{\pi\abs{\vec{k}}^2}{gh\omega^3}\Big\{(1+\cos\theta)\big[(\omega^2+f^2)\cos\theta-f^2\big]^2+\omega^2f^2[1-\cos(3\theta)]\Big\}{\hatt{E}(2\abs*{\vec{k}\sin(\theta/2)})},\label{isoscat}
\end{equation}
where we have removed the prime from $\theta'$ for convenience.
%where we have let $\sigma':=\int\sigma\abs{\vec{k}'}\d\abs{\vec{k}'}$. We note that this expression has no dependence on the direction of $\vec{k}$ because the scattering process is rotationally invariant. 
%where we have used the evenness of  $\sigma'$ in $\theta'$ to write the right-hand side as a convolution
Similarly, the total scattering cross section \eqref{Sigma} reduces to
\begin{equation}
    \Sigma(\abs{\vec{k}})=\int_{-\pi}^\pi\sigma'(\abs{\vec{k}},\theta)\d\theta. \label{Sigma_polar}
\end{equation}
In the limit $\omega \to f$ corresponding to near-inertial waves the cross section reduces to 
\begin{equation}
    \sigma'(\abs{\vec{k}},\theta) = \frac{2\pi f |\vec{k}|^2}{gh} \hatt{E}(2\abs*{\vec{k}\sin(\theta/2)}),
\end{equation}
which  recovers the result obtained by \citet{danioux} starting from the Young--Ben Jelloul model.

With the scattering cross section  \eqref{isoscat}, the scattering operator \eqref{La_polar} can be diagonalised using a Fourier series, or more precisely a cosine series since $a$ is even in $\theta$. Denoting  the cosine transform by a hat, with
\begin{equation}
\hatt{a}_n(\vec{x},|\vec{k}|,t)=\frac{1}{2\pi} \int_{-\pi}^{\pi} \cos( n \theta)  a(\vec{x},\abs{\vec{k}},\theta,t) \, \d \theta,   
\end{equation}
we find that
\begin{equation}
    (\widehat{\mathcal{L}a})_n=\lambda_n\hatt{a}_n,\;\;\;n=0,1,\cdots,
\end{equation}
with the eigenvalues 
\begin{equation}
    \lambda_n=\lambda_n(|\vec{k}|) :=2\pi\hatt{\sigma}'=\int_{-\pi}^{\pi}\sigma'(\abs{\vec{k}},\theta)\cos(n\theta)\d\theta.\label{lambda_n}
\end{equation}
Fourier transforming the kinetic equation \eqref{transport_equation} then gives
\begin{equation}
    \partial_t \hatt{a}_n + (\reallywidehat{\nabla_{\vec{k}} \omega \cdot \nabla_{\vec{x}} {a}})_n = (\lambda_n - \Sigma) \hatt{a}_n.
    \label{fourier_transport}
\end{equation}
It follows from \eqref{lambda_n} and the non-negativity of $\sigma'$ in \eqref{isoscat} that
\begin{equation}
    \lambda_0=\Sigma(\abs{\vec{k}})\;\;\;\text{and}\;\;\;\abs{\lambda_{n\geq 1}}<\lambda_0.\label{eigenvalue_properties}
\end{equation}
Thus the scattering term on the right-hand side of \eqref{fourier_transport} vanishes for $n=0$ and represents a damping for $n \ge 1$. 

The implications are clearly seen for a wave field that is spatially homogeneous, that is, with $\nabla_\vec{x} a=0$: the solution of \eqref{transport_equation}, with initial condition 
\begin{equation}
    a(\abs{\vec{k}},\theta,t=0)=A(\abs{\vec{k}},\theta)\label{initial_condition},
\end{equation}
is then simply
\begin{equation}
    a(\abs{\vec{k}},\theta,t)=\sum_{n=0}^\infty \hatt{A}_n(\abs{\vec{k}})\exp^{(\lambda_n-\Sigma)t}\cos(n\theta).\label{solution_reduced_transport_eqn}
\end{equation}
This describes the relaxation of the solution towards a stationary, isotropic ($\theta$-independent) solution, since
\begin{equation}
    \lim_{t\to\infty}a(\abs{\vec{k}},\theta,t)=\hatt{A}_0(\abs{\vec{k}})=\frac{1}{2\pi}\int_{-\pi}^{\pi}a(\abs{\vec{k}},\theta,t=0)\d\theta.
\end{equation}
This is a key feature of the scattering: the main impact of the random isotropic flow is to lead to the isotropisation of the IT field  regardless of the initial condition.Note that, with $a(|\vec{k}|,\theta,t)$ the wave-energy density, $\hatt{A}_0$ represents the total, $\theta$-integrated energy at wavevector $|\vec{k}|$, while the $\hatt{A}_n$ for $n \not= 0$ capture the energy's dependence on $\theta$.

We can identify two timescales for the scattering process. First, the scattering time
\begin{equation}
    T_{\mathrm{scat}}=\Sigma^{-1}\label{T_scat}
\end{equation}    
estimates the time over which energy concentrated at $\vec{k}$ in spectral space is reduced by a factor of $\exp^{-1}$ while  converted to waves with other wavevectors. In other words, it is the timescale over which scattering effects become significant. Second, the timescale for convergence to an isotropic wavefield is given by 
\begin{equation}
    T_{\textrm{iso}}=(\Sigma-\lambda')^{-1},\;\;\;\text{where}\;\;\;\lambda':=\max_{n\geq 1}\lambda_n.\label{T_iso}
\end{equation}
This is the time for the last surviving anisotropic (i.e.\ $n\neq 0$) Fourier mode to decay by a factor of $\exp^{-1}$.  Scattering lengthscales associated with the timescales \eqref{T_scat} and \eqref{T_iso} can be defined as
\begin{equation}
 L_{\mathrm{scat}}=c_g T_{\mathrm{scat}},\;\;\;L_{\mathrm{iso}}=c_g T_{\mathrm{iso}}  ,\label{lengthscales}
\end{equation}
where  $c_g=\abs{\nabla_\vec{k}\omega}=gh\abs{\vec{k}}/\omega(\abs{\vec{k}})$ is the group speed.

\subsection{Predicted behaviour} \label{sec:predicted}

In this section, we use the time- and lengthscales \eqref{T_scat}--\eqref{lengthscales} to examine how the scattering depends on the Coriolis parameter $f$, and on the strength and horizontal scales of the eddies as encoded in $\hatt{E}$. Since we focus on ITs, we regard the frequency $\omega$ as fixed and deduce $\abs{\vec{k}}$ from the dispersion relation \eqref{disp_relation}. We test some of our predictions against numerical simulations in \S\ref{sec:simulations}.

We assume an isotropic energy spectrum of the form
\begin{equation}
    \hatt{\mathcal{E}}(\abs{\vec{k}}):= 2\pi \,\abs{\vec{k}}\,\hatt{E}(\abs{\vec{k}}) =
    \begin{cases}
    c_1\abs{\vec{k}} & \qquad \abs{\vec{k}}\leq \kappa,\\
    c_2\abs{\vec{k}}^{-3.5} &\qquad \abs{\vec{k}}\geq\kappa.
    \end{cases}\label{iso_energy_spectrum}
\end{equation}
This depends on two parameters: $\kappa$, a peak wavenumber which sets the dominant lengthscale of the flow, and the root-mean-square velocity defined by $v_{\mathrm{rms}}^2 = \int_0^\infty\hatt{\mathcal{E}}\d\abs{\vec{k}}$. 
The constants $c_1$ and $c_2$ are determined by $\kappa$ and $v_{\mathrm{rms}}$ and the requirement of continuity at $|\vec{k}|=\kappa$. 
In practice, we choose $\kappa$ so that the correlation length $l_c:=\pi/k_c$, where $k_c:={\iint \abs{\vec{k}}\hatt{E}\d{\vec{k}}}/{\iint \hatt{E}\d{\vec{k}}}$, is similar to the wavelength of the IT; calculation of the integrals using \eqref{iso_energy_spectrum} gives $\kappa=9\pi/(10\, l_c)$.
Although quasigeostrophic theory predicts a kinetic energy spectrum that decays as $\abs{\vec{k}}^{-3}$ for balanced geostrophic turbulence, the slope is often observed to be slightly steeper, a result which is typically attributed to the presence of large-scale coherent structures that emerge in the turbulent flow \citep{mcwilliams,bartello,kafiabad}. This motivates the form of \eqref{iso_energy_spectrum} as representative of balanced geostrophic eddy fields in the ocean. Note that we have chosen an energy spectrum with non-zero energy for all $|\vec{k}|$. This matters because only the range  $[0, 2 |\vec{k}|]$ of the energy spectrum  contributes to the scattering of ITs with wavenumber $|\vec{k}|$, as the 
 factor $\hatt{E}(2\abs{\vec{k}\sin(\theta'/2)})$ in the scattering cross section \eqref{isoscat} indicates. A lower cutoff of the spectrum, say at some wavenumber $k_\mathrm{cut}$, would then imply that waves with $|\vec{k}|\le k_\mathrm{cut}/2$ are unaffected by scattering.
 This is related to the resonant triad view that two ITs with the same wavenumber $|\vec{k}|$ and hence the same frequency can only form a resonant triad with a flow mode if this has a wavenumber in $[0, 2 |\vec{k}|]$.
 
\begin{figure}
    \centering
    \begin{subfigure}[t]{.49\linewidth}
    \centering 
    %\caption{$\sigma'(\abs{\vec{k}},\theta')$ for $ \abs{\vec{k}}/2\leq \kappa \leq 2\abs{\vec{k}}$}
    \includegraphics[width=1\linewidth]{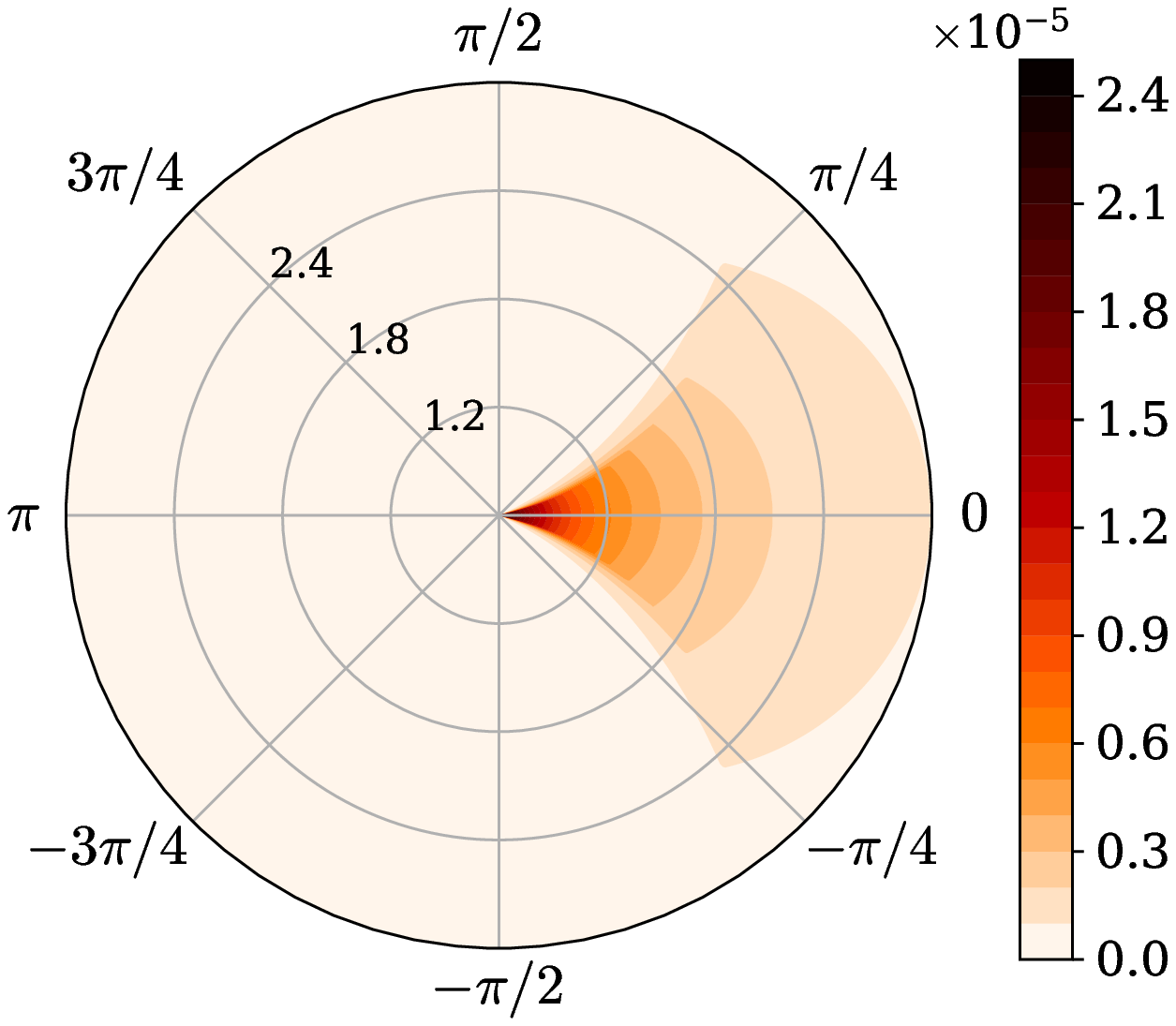}
        \label{fig:my_label}
    \end{subfigure}
    \begin{subfigure}[t]{.49\linewidth}
    \centering
     %  \caption{$\sigma'(\abs{\vec{k}},\theta')$ for $ \abs{\vec{k}}\leq k_c = 1.6\times 10^{-5}$}
        \includegraphics[width=1\linewidth]{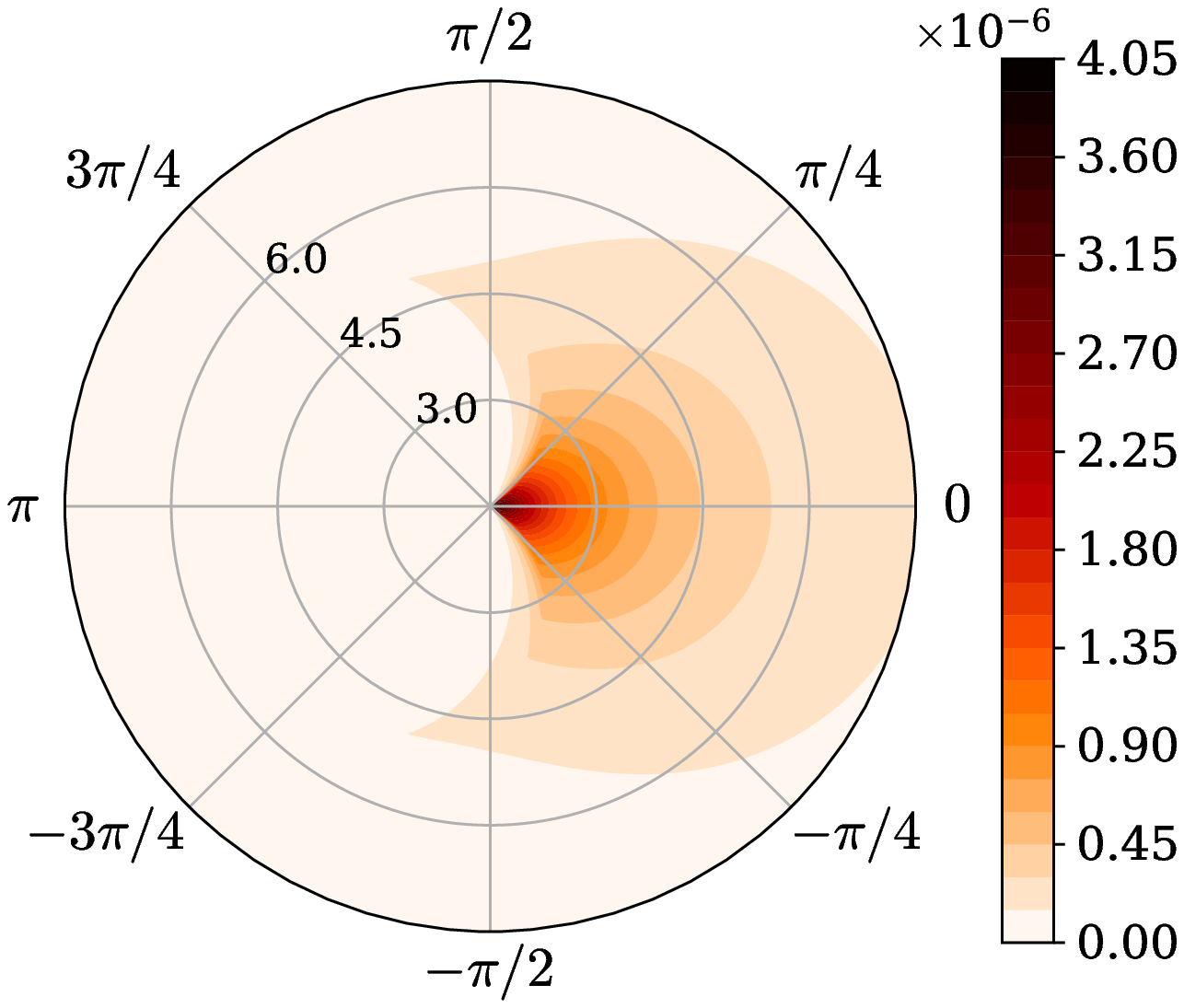}
    \end{subfigure}
     
    % \begin{subfigure}[t]{\textwidth}
    % \centering
    % \caption{$\sigma'(\abs{\vec{k}},\theta')$ for $ \abs{\vec{k}}\leq 2\times 10^{-6}$}
    %     \includegraphics[width=0.5\linewidth]{sigma3.eps}
    % \end{subfigure}
    \caption{Scattering cross section $\sigma'$ in \eqref{isoscat} for the energy spectrum \eqref{iso_energy_spectrum} as a function of the peak wavenumber $\kappa$ and  angle $\theta$ treated as polar coordinates. The IT wavenumber is fixed as $|\vec{k}|=3 \times 10^{-5} \, \textrm{m}^{-1}$ corresponding to the mode-$1$ $M_2$ tide at $45^\circ$ latitude, and the flow's root-mean-square velocity as $v_\mathrm{rms} = 0.25 \, \textrm{m} \, \textrm{s}^{-1}$. The left panel shows the range $0.2 |\vec{k}|\le \kappa \le |\vec{k}|$, the right panel the range $0.5 |\vec{k}|\le \kappa \le 2.5 |\vec{k}| $. Three (circular) contours of $\kappa$ are labelled in each panel in units of $10^{-5}$ m$^{-1}$.}\label{polar_plots}
\end{figure}

Figure \ref{polar_plots} shows the scattering cross section $\sigma'$ in \eqref{isoscat} as a function of $\theta$ and $\kappa$ for $v_\mathrm{rms}=0.25 \, \textrm{m} \, \textrm{s}^{-1}$ and for $f=1.028 \times 10^{-4} \, \textrm{s}^{-1}$ corresponding to 45$^\circ$ latitude. The equivalent depth is set to $h=1.2$ m, as appropriate for the first baroclinic mode \citep{olbers}, and the  frequency to  $\omega = 2\pi/12.42$ hours, corresponding to the $M_2$ tide.
The horizontal wavenumber  is then $|\vec{k}|=3 \times 10^{-5} \, \textrm{m}^{-1}$ corresponding to a wavelength of about $200 \, \textrm{km}$.
% ($\kappa=1.45\times 10^{-5}$ corresponds to a correlation length $l_c\sim 180$ km)
The figure indicates that scattering is local in the angular coordinate, that is, ITs are preferentially scattered into waves with nearby directions. This is especially the case for small values of $\kappa$, corresponding to flows with typical scales much larger than the IT wavelength (left panel), when the values of $\sigma'$ are also the largest. For larger values of $\kappa$, that is, for flows with smaller scales, the energy transfers are slower, but less localised in the angular direction (right panel).

\begin{figure}
\centering
\begin{subfigure}[t]{.46\textwidth}
    \centering
    % (a) \\
    \caption{ }
    \includegraphics[width=\linewidth]{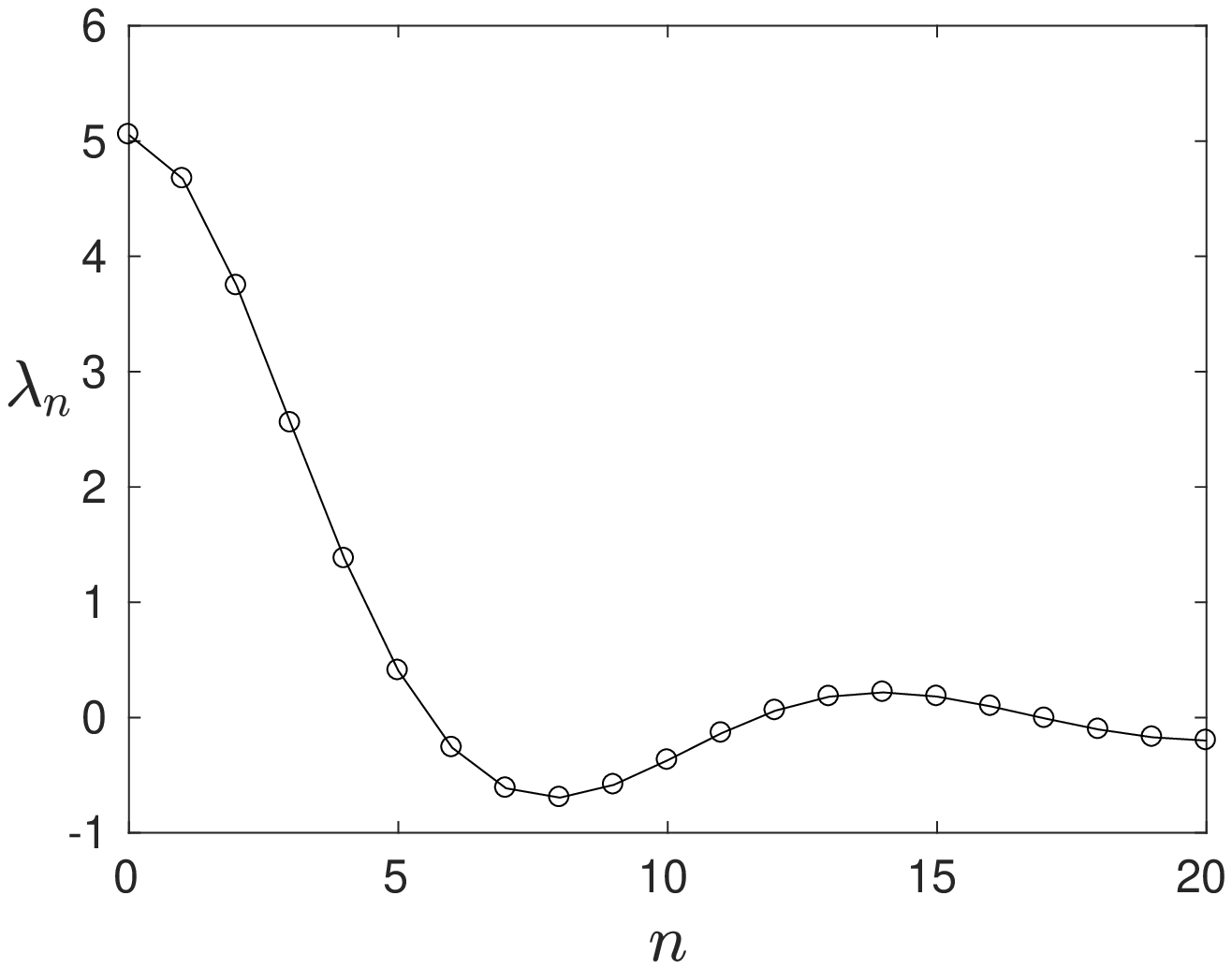}
    %\caption{Note to self - First 20 eigenvalues for energy spectrum with $\kappa$=1.15e-5, f=1.028e-4, v=0.24 }
    \label{fig4}
    \end{subfigure}\quad
\begin{subfigure}[t]{0.46\textwidth}
    \centering
    % (b) \\
    \caption{}
        \includegraphics[width=\linewidth]{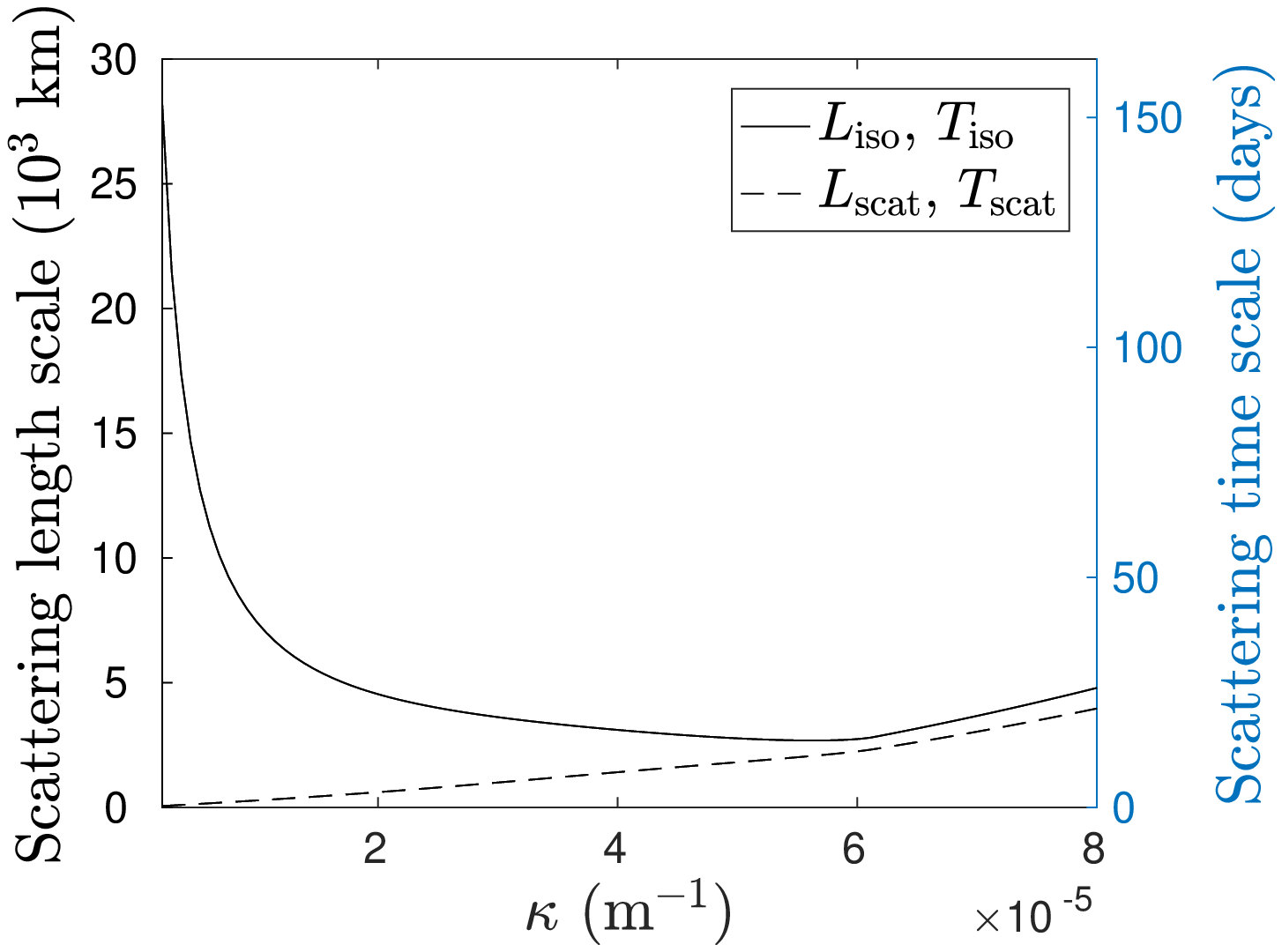}
    %\caption{Parameters -  $f=1.028\times 10^{-4}$ (latitude $45^\circ$), $v_{\mathrm{rms}}=0.25$m/s.}
    \label{fig1}
    \end{subfigure}

    \begin{subfigure}[t]{.46\textwidth}
     \centering
    %  (c) \\ 
    \caption{} 
    \includegraphics[width=\linewidth]{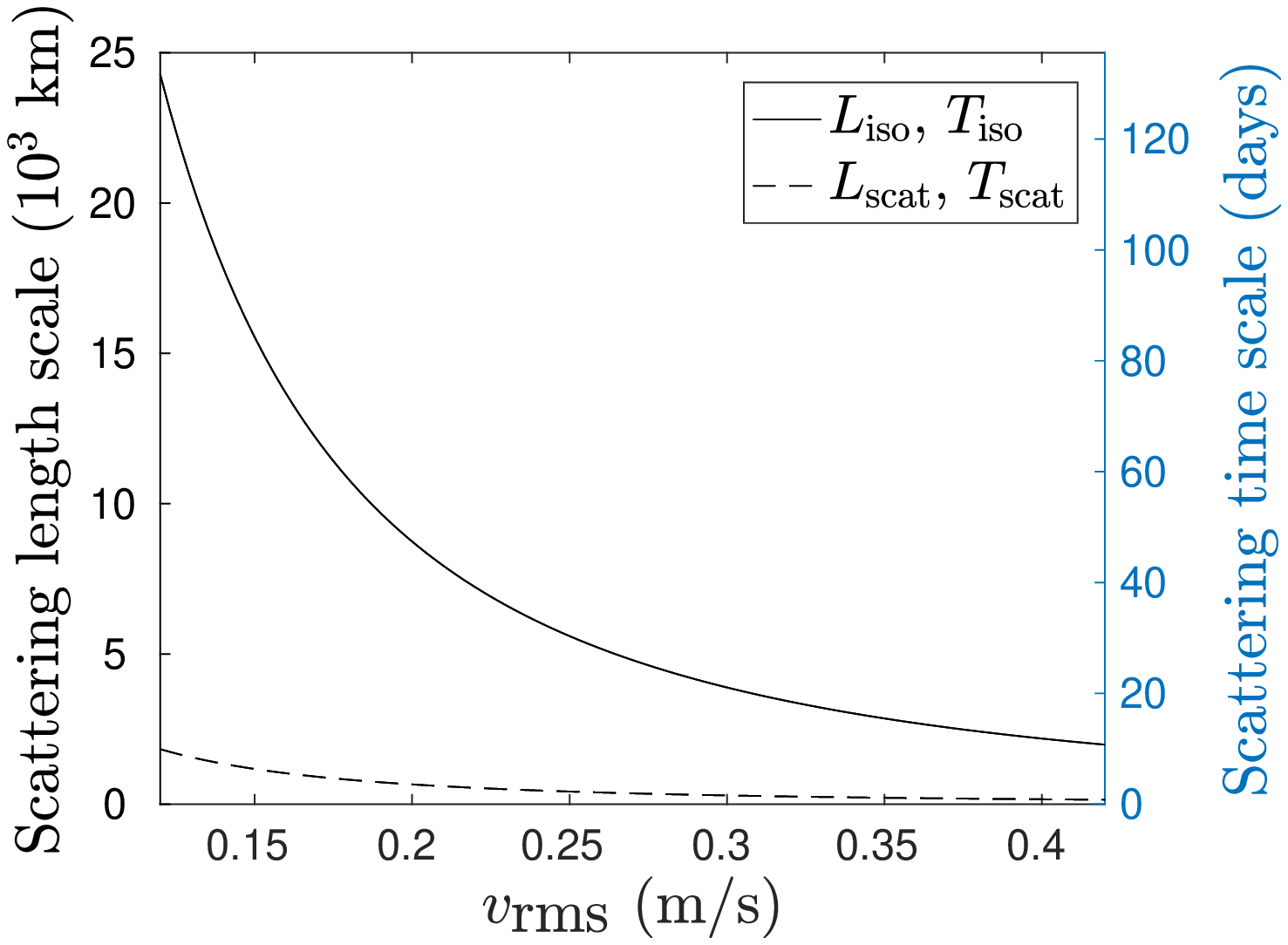}
    %\caption{Parameters - 45 degree north, f=1.028e-4, $\kappa$=1.45e-5 m-1, -3.5 spectrum \eqref{iso_energy_spectrum}  }
    \label{fig2}
    \end{subfigure}\qquad
    \begin{subfigure}[t]{.46\textwidth}
     \centering
    %  (d) \\
    \caption{ }
    \includegraphics[width=\linewidth]{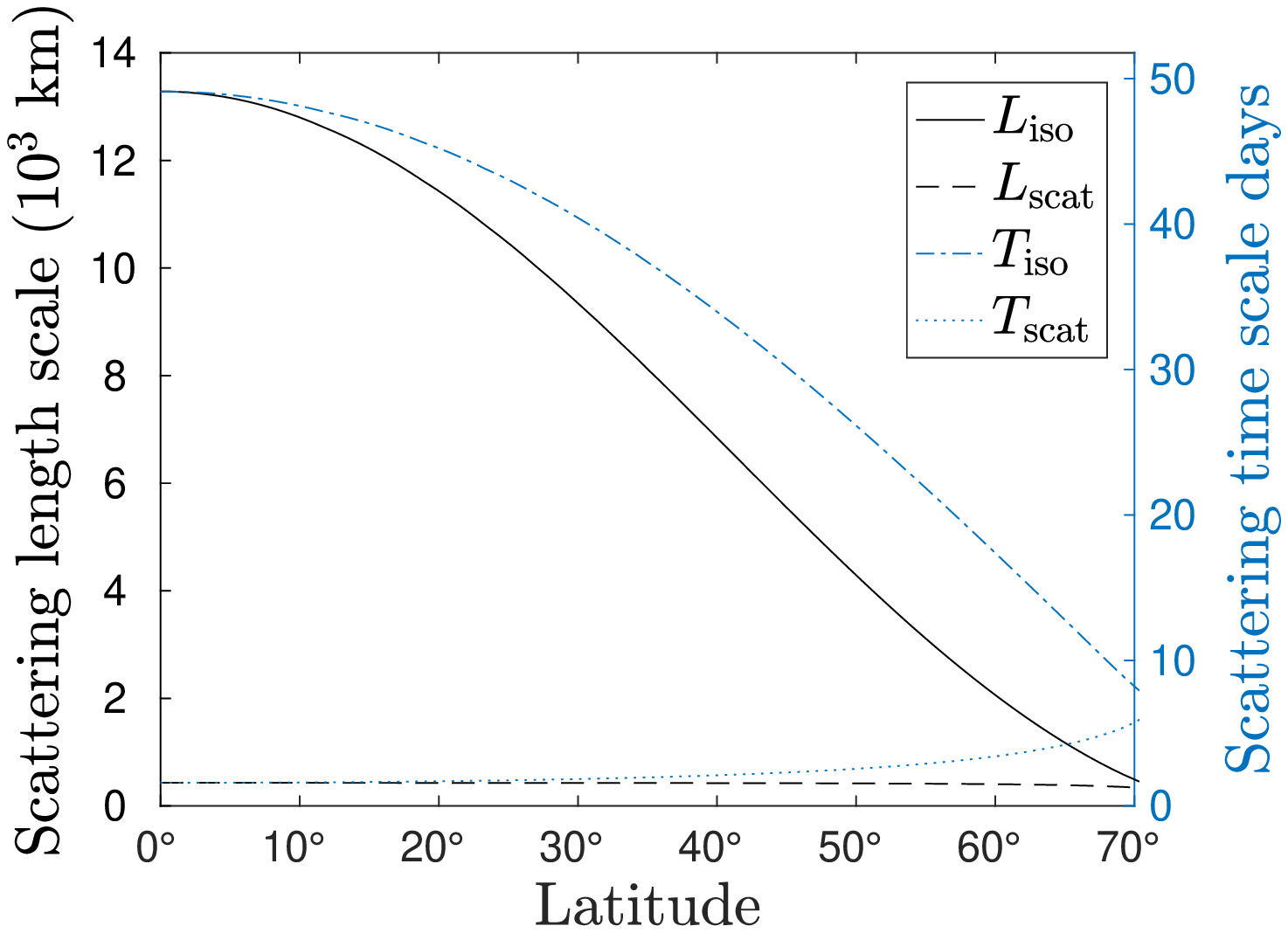}
    %\caption{ $\kappa$=1.45e-5 m-1, $v_{rms}=0.25$m/s}
    \label{fig3}
    \end{subfigure}\qquad
    
    \caption{(a) Eigenvalues $\lambda_n$ of the scattering operator, given by \eqref{lambda_n}, for the energy spectrum in \eqref{iso_energy_spectrum} with $\kappa=1.45 \times 10^{-5} \, \mathrm{m}^{-1}$ and $v_\mathrm{rms}=0.25 \, \mathrm{m} \, \mathrm{s}^{-1}$, and for an IT with $|\vec{k}|=3 \times 10^{-5} \, \mathrm{m}^{-1}$ and $f=1.028 \times 10^{-4} \, \mathrm{s}^{-1}$. (b) Scattering and isotropisation length and time scales $L_\textrm{scat}, \, L_\textrm{iso}, \, T_\textrm{scat}$ and $ T_\textrm{iso}$ as functions of the peak wavenumber $\kappa$, with all the other parameters as in (a). (c) As in (b) but as functions of $v_\mathrm{rms}$ and for $\kappa=1.45 \times 10^{-5} \, \mathrm{m}^{-1}$. (d) As in (b) but as functions of latitude.}
    \label{estimates}
    
    % (\ref{fig1} - \ref{fig3}) Length and time scale estimates for scattering by an isotropic random flow with an energy spectrum given by \eqref{iso_energy_spectrum}. An example of the eigenvalues obtained from \eqref{lambda_n} is shown in \ref{fig4}. The rate of isotropisation is determined by the gap between the two largest eigenvalues. }\
\end{figure}

The net effect of the scattering depends on both the value of $\sigma'$ at fixed $\theta$ and the range of $\theta$ where $\sigma'$ is substantial; it is best measured by the scattering and isotropisation time- and lengthscales introduced in \eqref{T_scat}--\eqref{lengthscales}. 
These scales are deduced from the cosine transform of $\sigma'$ which give the eigenvalues of the scattering operator. The eigenvalues are shown in Figure \ref{fig4}  
for $\kappa=1.45\times 10^{-5} \, \textrm{m}^{-1}$, corresponding to a flow correlation length $l_c\approx 180 \, \textrm{km}$, with all other parameters as in Figure \ref{polar_plots}. The most important eigenvalues are the two largest, $n=0$ and here $n=1$, since they control the scattering and isotropisation time- and lengthscales. 
These scales are displayed as functions of $\kappa$ in Figure \ref{fig1}. The figure shows that large-scale flows lead to rapid scattering but slow isotropisation. This can be easily understood: large-scale flows cause rapid energy transfers but, because of the localised nature of the scattering, these transfers are limited to waves of similar directions and a long time is needed for energy to be distributed near-uniformly in the angular direction. (The large-scale-flow regime can be tackled using ray tracing as has frequently been applied for deterministic flows \citep[e.g.][]{rainvillepinkel,chavanne}. For weak random flows as assumed here, the ray equations can be analysed asymptotically using methods developed for noisy Hamiltonian systems \citep[e.g.][]{bal-et-al} to show that the IT wavevector diffuses along the constant frequency circle $|\vec{k}|=\mathrm{const}$, consistent with the kinetic-equation description; see \citet{muller76,muller77} for early treatments in this spirit.)
Isotropisation is most effective when $\kappa$ has an order of magnitude similar to $|\vec{k}|$: for the chosen energy spectrum, isotropisation is fastest for $\kappa \approx 6 \times 10^{-5} \, \textrm{m}^{-1}$ corresponding to a flow correlation length of about $50 \, \textrm{km}$. Isotropisation slows down for larger values of $\kappa$ simply because the total flow energy in the useful range $[0,2|\vec{k}|]$ decreases with $\kappa$. 

% Above, we have plotted how the scattering time and length scales vary according to different parameters across typical oceanic ranges. The purpose of these is to give a sense of the functional dependence of the IT scattering on different variables, and to give concrete values that may be used to compare with real data in order to verify that the theory in the previous section may be used to make sensible predictions. The first figure \ref{fig1} gives an indication of how the scattering of the IT depends on the length scales of the eddies it propagates through. The parameter $\kappa$ controls the dominant wavenumbers of the background flow, such that smaller values of $\kappa$ give rise to larger scale eddies. By fixing the strength of the flow and the latitude, we observe that isotropisation takes place very slowly in very large scale flows. This isotropisation length scale rapidly decreases as the flow becomes comparable to the wavelength of the IT. {\color{blue}This dependence is not so surprising when considering the resonant triad interactions that scatter energy around the circle of constant frequency in spectral space.} If the flow is dominated by small wavenumbers, then the scattering mainly consists of small steps such that the difference between the initial and the scattered wave is small, and it takes a long time for the energy to distribute evenly into an isotropic state. If the flow is of a similar order to the wavelength of the IT, then energy can be scattered across the circle in larger jumps, leading to a more rapid onset of isotropy. 

Figure \ref{fig2} shows the scattering and isotropisation times and lengths as functions of $v_\mathrm{rms}$ and for 
$\kappa=1.45\times 10^{-5} \, \textrm{m}^{-1}$. The dependence is simply in $v_\mathrm{rms}^{-2}$. The figure suggests that full isotropisation of ITs generated at localised topographical features is rare in the ocean since the lengthscales required exceed the basin scales even for strong flows. On the other hand, scattering is effective over much shorter spatial scales, of the order of a few hundreds of kilometers, and over time scales of a week or so, comparable to other dynamical time scales in the ocean. 
The conclusion, then, is that typical ITs are strongly influenced by the quasigeostrophic flow, though not to the extent that they become completely isotropic.  
As highlighted by \citet{ward}, the timescale of a week or so is shorter than the characteristic timescales  of nonlinear wave--wave interactions except, perhaps, for the special case of parametric subharmonic instability at the critical latitude of $29^\circ$ \citep{mackinnon}. It is likely, then, that scattering by the geostrophic flow plays a more important role than wave--wave interactions in determining the characteristics of oceanic inertia-gravity waves. We should note, however, that the most energetic regions of the ocean, such as western boundary currents, where scattering is most effective, are also strongly inhomogeneous so that out theory does not apply in a strict sense.

Figure \ref{fig3} explores the dependence of scattering on latitude. Latitude affects the cross section $\sigma'$ through $f$ and also through $|\vec{k}|$ if we consider a fixed frequency as is done here. In the ocean, different latitudes may also lead to different energy spectra. For simplicity, in plotting  Figure \ref{fig3} we have taken the same spectrum for each latitude, keeping $\kappa$ fixed. 
The figure shows that the scattering time increases with latitude.
%as might be expected from the fact that the cross section decreases with $f$ and even vanishes for $f=\omega$. 
The isotropisation time, however, decreases with latitude with, as far as we can tell, no obvious interpretation; the scattering is determined as the difference between two eigenvalues and is hence difficult to intuit. The scattering and isotropisation lengths both decrease with latitude, partly as a result of a decrease of the group velocity.
We note that \citet{ward} conclude from simulations that scattering and isotropisation weaken with latitude, leading to longer propagation distances (see their Figure 11). This apparent contradiction is likely resolved by the fact that their non-dimensional formulation implies that their energy spectrum also changes with latitude, keeping the energy in the range $[0, 2 |\vec{k}|]$ constant as $f$ changes. A general conclusion  we can draw from the form of $\sigma'$ and our parameter-dependence study is the fact that the quasigeostrophic energy spectrum is the key factor determining the strength of the scattering.

\section{Simulations} \label{sec:simulations}
In this section we analyse numerical simulations of the linearised equivalent shallow-water system \eqref{SWEs} and compare them with the theoretical predictions of the previous section and with direct simulations of the kinetic equation \eqref{transport_equation}.

\subsection{Shallow-water simulations}

We solve \eqref{SWEs} numerically, adding a harmonic forcing term to generate a coherent plane wave. The numerical scheme relies on pseudospectral  and splitting methods: the terms independent of the background flow are integrated exactly in Fourier space, while the terms that depend on the background flow are integrated using an Euler scheme in physical space. The domain is a 7168 km $\times$ 1024 km channel on an $f$-plane centred at $45\degree$N. %where $f=1.028\times 10^{-4}$. 
We use a spatial resolution of  $1792 \times 256$, with that $\Delta x=\Delta y=4$ km, with periodic boundary conditions in the $y$-direction, and absorbing layers 30-gridpoints wide at each end of the domain in the $x$-direction, and take timesteps of $\Delta t = 4000$ s. We run an ensemble of 100 simulations with random realisations of the background flow in order to study statistics. Each simulation corresponds to 80 days, which is long enough to study isotropisation in a moderately energetic flow. 
%The equivalent depth is set to $h=1.2$ m, as appropriate for the first baroclinic mode \citep{olbers}, and the forcing frequency to  $\Omega = 2\pi/12.42$ hours, corresponding to the the $M_2$ tide. 
A wavemaker forces an IT through a term of the form  
\begin{equation}
\mathcal{F}=A \sin (\Omega t)\exp^{-(x-x_0)^2/\Delta^2},
\end{equation}
added to the continuity equation \citep[cf.][]{ponte}. Here, $x_0=400$ km is the position of the wavemaker in the $x$-direction and  $\Delta=10$ km its width; $\Omega$ is the tidal frequency. The amplitude $A$ is arbitrary since we solve a linear system.
The forcing is ramped up slowly to reach its maximum amplitude over approximately 1 week. The resulting plane waves that are generated have a wavelength of approximately $150$ km, as expected for a first baroclinic mode wave at $45^\circ$ latitude. 

\begin{figure}
\centering
\hspace*{-0.3cm}\includegraphics[width=1\linewidth]{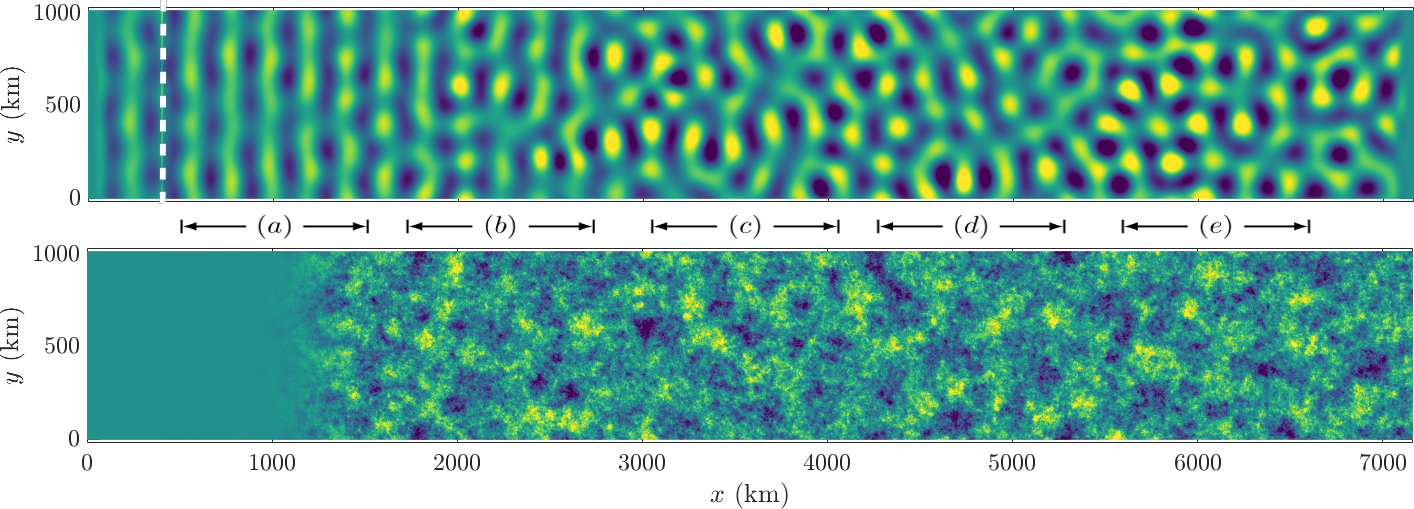}%\hspace*{-2.3cm}\includegraphics[width=1.3\linewidth]{results1viridis.eps}
    \centering 
    \caption{Top: sea-surface elevation $\eta$ in an equivalent shallow-water simulation of the mode-$1$ $M_2$ tide in a turbulent flow with $v_\textrm{rms}=0.25\, \textrm{m}\, \textrm{s}^{-1}$ at $45^\circ$ latitude. Bottom: vorticity field of the turbulent flow.}
    \label{fig5}
\end{figure}

For the background flow we take a homogeneous isotropic Gaussian random field, tapered in the region $0<x<1000$ km so as not to interfere with the wavemaker. This field is generated numerically as a Fourier series with random coefficients. The  energy spectrum is that in \eqref{iso_energy_spectrum}, with $v_{\mathrm{rms}}=0.25 \, \mathrm{m s}^{-1}$ and $\kappa=1.45\times 10^{-5}\, \mathrm{m}^{-1}$ leading to flows with a correlation length of about 200 km. 
The other parameters are those of the mode-1 $M_2$ tide at $45^\circ$, as in \S\ref{sec:predicted}.
The results presented below use a time-independent flow. We carried out additional simulations with slowly time-varying flows to confirm the theoretical prediction of the Appendix that  IT scattering is essentially unaffected by the time dependence of stationary random flows with timescales  $O(\Ro ^{-1})$ longer than the IT period. Note that the modelling of the background flow by a Gaussian random field is a choice motivated by practicality and the fact that, according to our theory, the only statistical property of the flow that influences the scattering is its energy spectrum. An alternative would be to carry out a large number of quasigeostrophic simulations to generate an ensemble of flows with more realistic statistics. This would however be computationally expensive; it would also require great care to control the flow parameters and to ensure stationary statistics.

\begin{figure}
     \centering
    %\hspace*{-1.5cm}
    \includegraphics[width=1\linewidth]{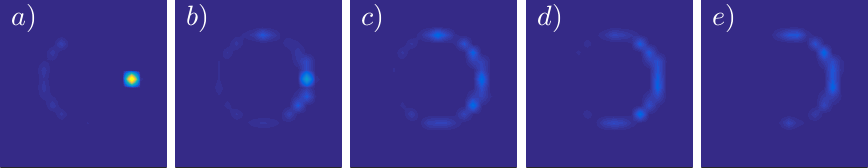}
    \caption{Energy density in the $\vec{k}$ plane of a single mode in different regions along the channel, from an ensemble of 100 shallow-water simulations. The regions are five {1024 km $\times$ 1024 km} boxes centred about the midpoints $x=\{1000,2250,3500,4750,6000\}$ km.}
    \label{fig6}
\end{figure}

 The top panel of Figure \ref{fig5}  shows one realisation of the IT height field $\eta$ at $t=80$ days, when plane waves generated by the wavemaker (indicated by a dashed line) have propagated across the eddy field shown in the bottom panel. Close to the wavemaker the wave field has a plane-wave structure which becomes scrambled in appearance as the phases randomise due to flow scattering. In agreement with our scattering theory, the wave field retains a single lengthscale -- the wavelength set by the tidal frequency -- throughout its evolution: scattering does not lead to a scale cascade. This is consistent with the earlier simulation results of \citet{ward},  \citet{wagner_ferrando_young_2017}, \citet{ponte} and \citet{dunphy}, the latter two in a three-dimensional setup. Note that, 
 since we solve the linearised system, there is no harmonic generation and consequent formation of smaller wave scales as described by \cite{ward}. 
 The eigenvalues $\lambda_n$ for the IT and flow parameters chosen are those shown in Figure \ref{estimates}(a) and correspond to $L_\textrm{scat}=420 \, \textrm{km}$ and $L_\textrm{iso}=5,600 \, \textrm{km}$. This is qualitatively consistent with the wave field in  Figure \ref{fig5}.
 
To assess our theoretical results in more detail, we need to estimate the energy density $a(\vec{x},\vec{k},t)$ from the simulations. To this end, we take Fourier transforms of the wave fields in five 1024 km $\times$ 1024 km square boxes spanning the length of the domain. For each realisation of the flow, we compute the Fourier transform of  $u,\,v$ and $\eta$ in each box at the end of the simulation, project onto the IT eigenmode then average over the ensemble to obtain an approximation $\tilde a(\vec{k},t)$, say, of $a(\vec{x},\vec{k},t)$ in the box. The details of this are given in Appendix \ref{app:modes}.
%  for each realisation we take the FFT of each wave field in 1024 km $\times$ 1024 km boxes positioned at midpoints along the channel, and then ensemble average the result for each box {\color{blue}(using these windows in space recovers some of the `$x$ dependence' of our sampled Wigner?)}. These are then used to calculate the spectral energy density of the waves in the neighbourhood about the midpoints. In Appendix  we also show that the wave fields can be decomposed into contributions from the different branches of the dispersion relation, so that we can track the evolution of just the $\omega_+$ mode via a projection onto its corresponding eigenvector. 
 The results are shown in Figure \ref{fig6}. For the first box, located immediately to the right of the wavemaker,  most of the energy is concentrated at the single point $\vec{k}=(k_0,0)$, where $k_0=\sqrt{(\omega^2-f^2)/gh}$, indicating a pure plane wave propagating to the right. (There is also a faint signal of left-propagating waves resulting from scattering at larger $x$.) In the next boxes, energy spreads around the circle of constant radius $\abs{\vec{k}}=k_0$. The spectrum is in fact distributed over a finite-width annulus rather than a circle, as a result of off-resonant interactions between waves and flow.
 
\begin{figure}
%\hspace{-0.25cm}%\centering
\begin{subfigure}[t]{.5\textwidth}
    \centering
    % (a) \\
    \caption{ }
    \includegraphics[width=\linewidth]{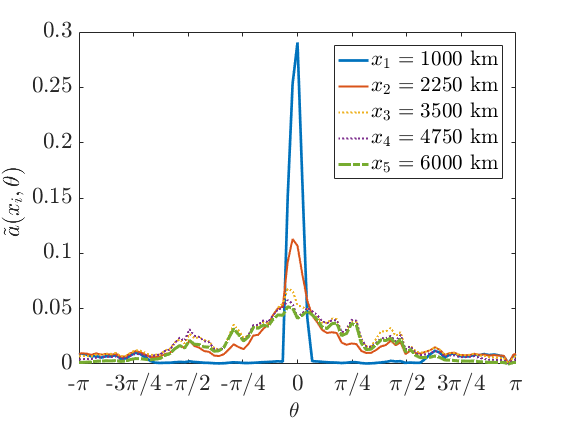}
    \label{fig7}
    \end{subfigure}
    \begin{subfigure}[t]{.5\textwidth}
     \centering
    %  (b) \\
    \caption{ }
    \includegraphics[width=\linewidth]{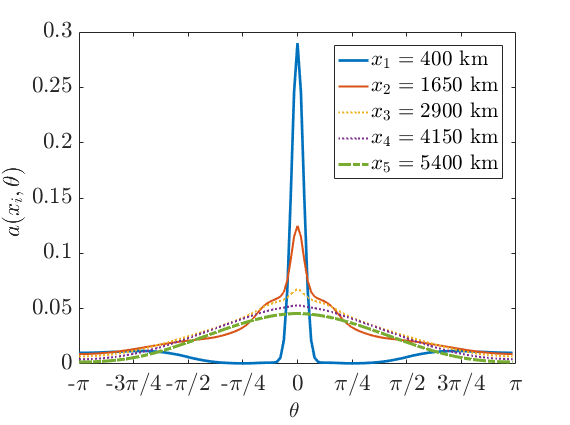}
   % \caption{PLACEHOLDER FIGURE - NB the offset is in part due to the fact that there is no random flow until 1000km, and so the data needs to be shifted ~5-600 km. Replot this figure with that in mind.}
    \label{fig8}
    \end{subfigure}
    \caption{Estimated energy density $\tilde a$  as a function of $\theta$ from (a) the ensemble of shallow-water simulations, with the $x_i$ taken as the box midpoints, and (b) the kinetic-equation simulation. The parameters are those of Figure \ref{fig5} and the time corresponds to the end of the simulation.}\label{shallow_vs_wigner}
\end{figure}

 We obtain a clearer view of the distribution of energy as a function of $\theta$ by integrating $a(\vec{x},\vec{k},t)$ in the 90 angular sectors $2(n-1)\pi/90 \le \theta \le 2n\pi/90,\, n=1,\cdots,90$. The results are shown in Figure \ref{fig7}. They enable a better assessment of the validity of the lengthscale estimates $L_{\mathrm{scat}}=420\, \textrm{km}$ and $L_{\mathrm{iso}}=5600 \, \textrm{km}$. Note that these lengths should be measured from the point where the background flow starts, which is at approximately $x=1000$ km, so that we would expect to see the fields isotropise at $x>6000$ km along the channel, corresponding to the rightmost box of Figure \ref{fig6}. We see however that the field is not fully isotropic in that region. An explanation is that the numerical simulation includes an absorbing layer near the right boundary of the domain, which allows energy to exit the channel but not to re-enter it. As a result, there are no left-propagating waves at the end of the channel. In addition, we emphasise that $L_{\mathrm{iso}}$ is only an order-of-magnitude estimate which, by converting timescale into lengthscale using the group speed, ignores the  directional properties of the transport of wave energy with the group velocity. We next go beyond this order-of-magnitude estimate and make direct predictions for the scattering by solving the kinetic equation numerically.

%  ***Talk more about the boundary conditions we've taken, and how by including absorbing layers, the assumption of a spatially homogeneous wave field does not hold, and so the predictions for the scattering lengthscales won't be entirely applicable. (Can we argue that they are still useful as a rule of thumb?) We can avoid the complexities of analytically dealing with the boundaries etc by simulating the kinetic equation \eqref{transport_equation} subject to the same boundary conditions. ***
 
 \subsection{Kinetic equation simulations}
 
 We simulate the kinetic equation \eqref{transport_equation} under the assumption of homogeneity in the $y$-direction, consistent with the periodic boundary conditions, and using the angle $\theta$, with $\vec{k}= k_0(\cos \theta, \sin \theta)$, as an independent variable. This reduces the number of independent variables  to 3, with $a(x,\theta,t)$, so that the kinetic equation becomes
 \begin{equation}
 \partial_t a+c_g\cos\theta \, \partial_x a=(\mathcal{L}-\Sigma) a +\mathcal{F}(x,\theta),\label{reduced_transport_eqn}
 \end{equation}
 where $\mathcal{L}$ and $\Sigma$ are given by \eqref{La_polar} and \eqref{Sigma_polar},
 %$c_g=\abs{\nabla_{\vec{k}}\omega}$ is the group speed, 
 and  $\mathcal{F}(x,\theta)$ is a forcing term mimicking the wavemaker of the shallow-water simulations. 
 We take $\mathcal{F}$ to be a Gaussian centred about $x=400$ km, $\theta=0$, with width parameters $\Delta_x=40$ km and $\Delta_\theta=0.1$, and an amplitude that is scaled to match the initial energy peak from the shallow-water simulation data.

\begin{figure}
     \centering
    %\hspace*{-1.5cm}
    \includegraphics[width=0.95\linewidth]{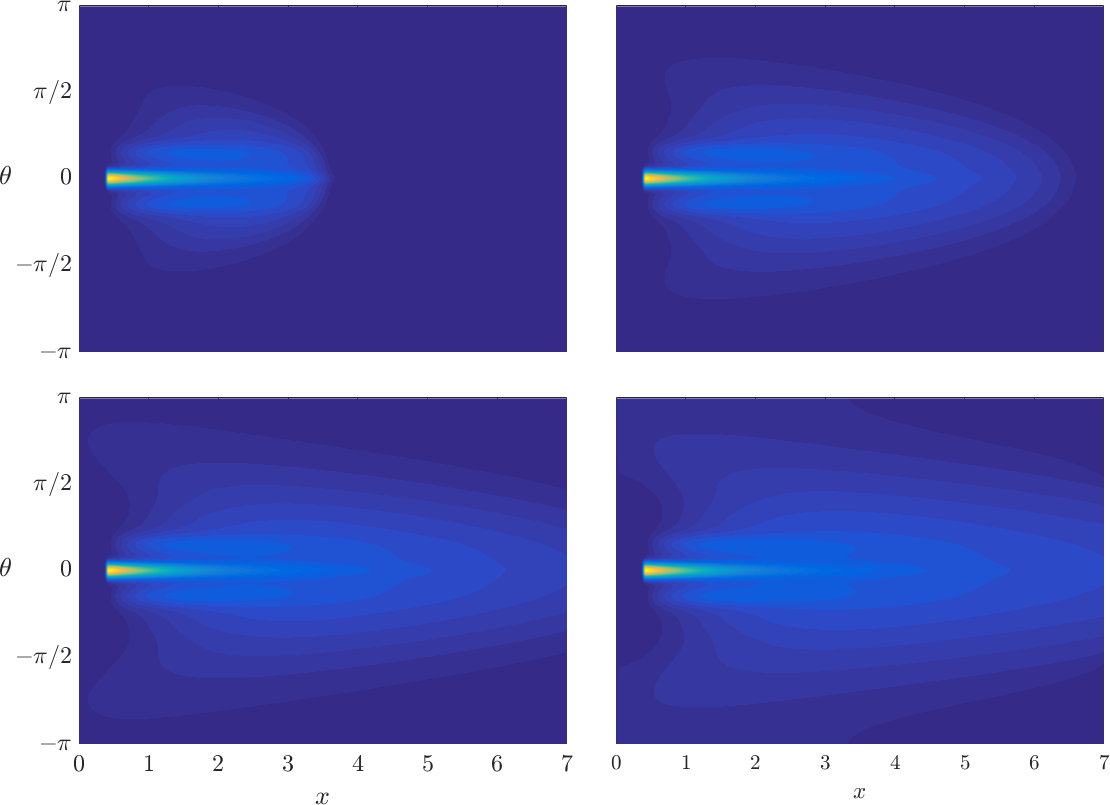}\qquad\qquad
    \caption{Wave-energy density $a(x,\theta,t)$ at $t=8, \, 16, \, 32$ and $80$  days obtained by solving the kinetic equations numerically with parameters matching those of Figure \ref{fig5}. The $x$-axis is in units of 1000 km.}
    
    % Note to self - taken at times $t=$8, 16, 32 and 80 days (last one corresponds to the end time of the other figures). $x$-axis is in 1000s of kms. This will all be tidied up...\\
    % Gaussian forcing located at $(\theta,x)=(0,400\textrm{km})$, with narrow width parameters.}
    \label{fig9}
\end{figure}

 We simulate \eqref{reduced_transport_eqn} using a pseudospectral splitting method. The advection term is integrated using a semi-Lagrangian finite-difference scheme, and the scattering and forcing terms  on the right-hand side are integrated exactly in Fourier space. The domain is 7168 km $\times$ $2\pi$, with a resolution of 1792 $\times$ 256, and time step $\Delta t= 900$ s up to a final time of 80 days. We apply periodic boundary conditions in the $\theta$-direction, and place absorbing layers 30-gridpoints wide at each end of the domain in the $x$-direction, as in the shallow-water simulation. The evolution of $a(x,\theta,t)$ is illustrated in Figure \ref{fig9}. The wave energy, initially concentrated at $(x,\theta)=(400,0)$, gets advected by the group velocity in the $x$-direction and spreads in the angular direction. Once energy reaches $|\theta|>\pi/2$, it propagates to the left, leading to the weak signal for $k<0$ observed in the first panel of Figure \ref{fig6}. 
 
 At the end of the simulation, we evaluate $a(x,\theta,t)$ at different points along the channel to get a set of curves $a(x_i,\theta,t=80 \,\mathrm{days})$, $1\leq i \leq 5$. The points $x_i$ are taken to be spaced along the channel in the same way as the windows shown in Figure \ref{fig6}. Note that we have to take into account the fact that there is no flow in the shallow-water simulations from the point of generation at $x=400$ km to approximately $x=1000$ km (see Figure \ref{fig5}), whereas the kinetic equation assumes a flow is present throughout the domain. To resolve the discrepancy, the values of $x_i$ are taken $600$ km less than the midpoints of the boxes used for the shallow-water simulations. The functions $a(x_i,\theta)$ are shown next to the equivalent shallow-water estimates in Figure \ref{shallow_vs_wigner}. There is a remarkably good agreement between the solution of the kinetic equation and the shallow-water simulation results. This demonstrates the value of the kinetic equation in predicting the generic properties of the scattering and their dependence on the various parameters in the problem.

\begin{figure}
%\hspace{-0.25cm}%\centering
\begin{subfigure}[t]{.5\textwidth}
    %\centering
    \caption{ }
    \includegraphics[width=\linewidth]{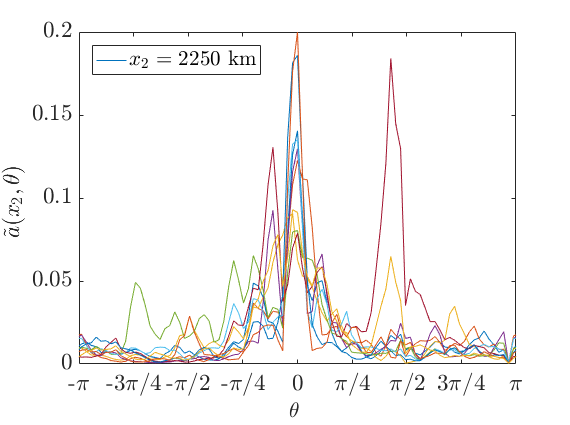}
        %\label{fig7}
    \end{subfigure}
    \begin{subfigure}[t]{.5\textwidth}
     %\centering
    \caption{ }
    \includegraphics[width=\linewidth]{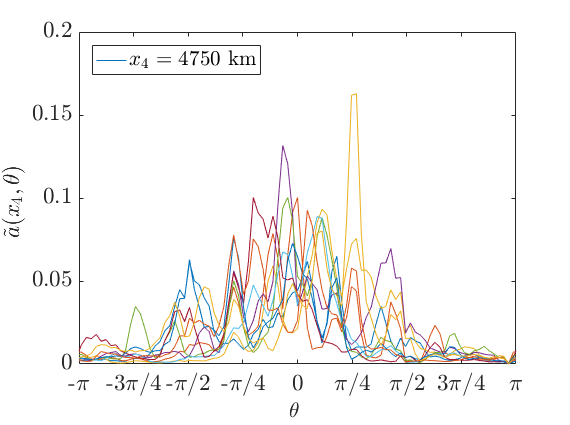}
    %\label{fig8}
    \end{subfigure}
    \caption{Energy density from 10 realisations of shallow-water simulations evaluated (a) in the second box of Figure \ref{fig6}), centred about $x_2=2250$ km, and (b) in the fourth box, centred about $x_4=4750$ km.}
    \label{fig10}
    %\caption{Comparison of Wigner function at $t=80$ days at a selection of distances along the channel with shallow-water simulations averaged over an ensemble of 100 runs. ***creat another figure to show how the data appears for 2-3 individual simulation for comparison, to show variability***}
\end{figure}
 
 We emphasise that the kinetic equation yields only ensemble average predictions and cannot describe the details of the effect of a single flow realisation on the IT. To illustrate how the scattering fluctuates between realisations, we show in Figure \ref{fig10} the energy density $\tilde a$ estimated from single shallow-water simulations. While fluctuations can be large, the typical behaviour is a redistribution of energy in the angular direction that is well captured by the ensemble-averaged predictions. Furthermore, ergodicity  implies that the ensemble average deductions of the kinetic equation apply accurately to quantities that are spatial averages over many eddy scales.

% \FloatBarrier
% \iffalse
% \FloatBarrier

% \begin{figure}
%      \centering
%     %\hspace*{-1.5cm}
%     \includegraphics[width=0.9\linewidth]{Wigner_powerlaw_simulation2.png}
%     \caption{Note to self - simulation after 64 days with -3.5 spectrum, 1.6e-5 cutoff, 512x256 resolution, absorbing layers $\sim$ 50km at sides,}
%     \label{fig4}
% \end{figure}
% \fi

\section{Discussion} \label{sec:discussion}

This paper examines the scattering of oceanic ITs -- or indeed of any inertia-gravity wave -- caused by the turbulent mesoscale flow in which they propagate. Assuming that the flow is barotropic, weak (small Rossby number) and random with stationary and homogeneous statistics, we derive the kinetic equation \eqref{transport_equation} governing energy exchanges between waves travelling in different directions. A key outcome is the scattering cross section \eqref{cross_section}, or \eqref{isoscat} for an isotropic flow, which measures the energy transfer rate as a function of the wavevectors involved and other parameters. The scattering cross section depends linearly on the energy spectrum of the flow.

The form of the scattering cross section shows that the energy exchanges between waves are restricted to waves with the same frequency and hence the same horizontal wavenumber. Therefore, while scattering results in a complex random wavefield, this field has a single spatial scale determined by the forcing frequency. This is obvious for a time-independent flow but perhaps less so for the time-dependent flows we consider for which it is a consequence of the statistical stationarity of the flows. Note that this does not imply that the wavefield is completely phase-locked in time: slow phase variations result from the interaction with a time-dependent flow \citep{ponte,dunphy}, but these are not described by our analysis, which focuses on the wave amplitude as measured by the energy density $a(\vec{x},\vec{k},t)$. It would be of interest to study the phase variations from the statistical viewpoint taken here. 

For an isotropic flow, scattering leads to an equilibrium isotropic wavefield over a time scale that we can estimate from the Fourier transform of the scattering cross section \eqref{isoscat}. At equilibrium, and in the absence of spatial modulations, the wave energy density is $a(\vec{k}) = a(|\vec{k}|) \propto  \delta(|\vec{k}|- k_0)$ with $k_0=\sqrt{(\omega^2-f^2)/gh}$, corresponding to a correlation function $\int a(|\vec{k}|) \exp^{\i \vec{k} \cdot \vec{x}} \, \d \vec{k} \propto  \int \delta(|\vec{k}|-k_0) \exp^{\i \vec{k} \cdot \vec{x}} \, \d \vec{k}  $ that is 
proportional to the Bessel function $J_0(k_0 |\vec{x}|)$. Thus, while the flow 
controls the speed of convergence towards the isotropic wave-energy distribution, it has no effect on the form of this distribution. 

The time scale necessary for scattering to significantly alter the wave field is deduced from the scattering cross section and found to be of the order of a few days to a week. This is short compared with time scales associated with nonlinear wave--wave interactions \citep[see][]{ward}, which raises the possibility that scattering is as crucial as the more widely considered wave--wave interactions in shaping the inertia-gravity-wave spectrum in the ocean. Note however that, as our asymptotic treatment makes clear, the large time-scale separation between inertia-gravity waves and the quasigeostrophic flow implies that scattering causes little frequency broadening and so cannot by itself explain the continuum of observed frequencies.

The conclusion that scattering simply relaxes wave energy towards an isotropic equilibrium depends crucially on the two-dimensional setup implied by our assumption of barotropic flow. It holds because scattering redistributes wave energy in Fourier space over constant-frequency curves which, in this case, are just circles, hence compact. This picture changes radically in the presence of vertical shear since this causes energy exchanges between different vertical wavenumbers. The relevant constant-frequency set is then a cone in wavevector space. Because this cone is unbounded, no finite-energy equilibrium exists, and the energy of an initially plane wave can be expected to cascade to small scales, both horizontally and vertically and with a fixed aspect ratio, as it spreads on the cone. This scenario, already envisioned by \cite{lelong} and \cite{bartello}, is potentially important for the dissipation of oceanic inertia-gravity waves, with implications for their impact on mixing and mesoscale dynamics and for the maintenance of balance. The framework we have adopted, which regards the flow as a prescribed random field and examines the wave statistics on the basis of a kinetic equation, generalises to the case of vertically sheared flow. It is well suited to describe and quantify the scale cascade that results from what is then a fully three-dimensional scattering. This is the subject of future work. \medskip

\noindent
\textbf{Acknowledgements.} This research is funded by the UK Natural Environment Research Council under the NSFGEO-NERC programme (grant NE/R006652/1). M.A.C.S was supported by The Maxwell Institute Graduate School in Analysis and its Applications, a Centre for Doctoral Training funded by the UK Engineering and Physical Sciences Research Council (grant EP/L016508/01), the Scottish Funding Council, Heriot-Watt University and the University of Edinburgh.

\appendix
\section{Vertical-mode decomposition}\label{app:vmode-decomp}
Since the background flow in \eqref{boussinesq1} is barotropic, we project onto a vertical-mode basis according to 
\begin{equation}
    \left(\begin{array}{c}
        u\\
        v\\
        w\\
        p\\
        b
    \end{array}\right)=\sum_{m=0}^\infty\left(\begin{array}{cc}
       u_m(x,y,t) & F_m(z) \\
       v_m(x,y,t) & F_m(z) \\
       w_m(x,y,t) & N^{-2}(z) F'_m\\
       p_m(x,y,t) & F_m(z) \\
       b_m(x,y,t) & F'_m
    \end{array}\right),\label{vmodeansatz}
\end{equation}
{where the $F_m$ are eigenfunctions of the Sturm--Liouville problem }
\begin{equation}
   {\mathscr{L}F_m=-\frac{1}{r_m^2}F_m, \quad F_m'(0)+N^2 F_m(0)/g=0, \ \ F_m'(-H)=0, }
\end{equation}
{where $\mathscr{L}(\cdot)=\partial/\partial z[f^2/N^2\partial/\partial z(\cdot)]$ and $H$ is the ocean depth \citep{olbers}. The eigenvalues $r_m$ are the Rossby radii of deformation. Orthonormality implies}
\begin{equation}
  { (F_n,F_m)=\int_{-H}^0F_mF_n\d z=\delta_{mn}.}
\end{equation}

{Substituting \eqref{vmodeansatz} into \eqref{boussinesq1} leads to a system for the amplitudes $u_m, \, v_m$, etc.\ of each baroclinic mode $m$.
Defining the equivalent depth $h_m=f^2 r_m^2/g$ and the equivalent surface height $\eta_m=b_m/g$, we rewrite this system in the shallow-water-like form given by \eqref{SWEs}.}
\section{Kinetic equation derivation} \label{app:kinetic}

In this appendix we apply the formalism of \cite{ryzhik} to derive a kinetic equation for the equivalent shallow-water system \eqref{SWEs} (see also \citet{bal-et-al} and \citet{powell}). This formalism exploits the weakness of the flow as measured by the small Rossby number, as well as the scale separation between wavelength and flow scale on the one hand, and the scale of variation of the wave amplitude on the other hand. The associated small parameter, $\varepsilon$, is assumed to satisfy $\varepsilon=O(\Ro^2)$.

\subsection{Shallow-water equations and scaling regime}

We start by rewriting \eqref{SWEs} in the more abstract notation
\begin{equation}
    \partial_t\vec{\phi}+\big[L(\partial_\vec{x})+\sqrt{\varepsilon}N({\vec{x}},\partial_\vec{x},\sqrt{\varepsilon}t)\big]\vec{\phi}=0,\label{dynamics}
\end{equation}
where the vector
\begin{equation}
    \vec{\phi}(\vec{x},t)=(u,\,v,\,\eta)^\mathrm{T}
\end{equation}
groups the dynamical variables and we have introduced the matrix operators
\begin{equation}\arraycolsep=1.6pt
    L(\partial_x)=\left(\begin{array}{ccc}
        0 & -f & g\partial_x\\
         f & 0 & g\partial_y\\
         h\partial_x & h\partial_y &0
    \end{array}\right) \quad \textrm{and} \label{L_operator}
\end{equation}
\begin{equation}
        \arraycolsep=0.2pt N({\vec{x}},\partial_\vec{x},\sqrt{\varepsilon}t)=\left(\begin{array}{ccc}
        \psi_x\partial_y-\psi_y\partial_x-\psi_{xy} & -\psi_{yy} &0 \\
        \psi_{xx} & \psi_x\partial_y-\psi_y\partial_x+\psi_{xy}&0\\
        0&0&\psi_x\partial_y-\psi_y\partial_x
    \end{array}\right). \label{N_operator}
\end{equation}
In \eqref{dynamics}, we have made the relative importance of the various terms explicit by scaling them with the relevant power of the small parameter $\varepsilon$. For convenience, we keep the equations in their dimensional form and treat $\varepsilon$ as a bookkeeping parameter that can be set to $1$ at the end of the calculation. Note that we could use $\Ro$ as an alternative to $\varepsilon$ in this bookeeping role and, in this manner, avoid fractional powers; using $\varepsilon$ keeps our derivation in closer correspondence to that of \cite{ryzhik}. Note also that the dependence of $N$ on $\sqrt{\varepsilon}t$ arises through the slow time dependence of $\psi$.

The depth-averaged energy density for the shallow-water system without background flow is given by
\begin{equation}
   \mathcal{E}(\vec{x},t)=\tfrac{1}{2}\left(h(u^2+v^2)+g\eta^2\right) = \tfrac{1}{2}\inner{\vec{\phi}}{\vec{\phi}}_M,
\end{equation}
where we have defined the inner product
\begin{equation}
 \inner{\vec{f}}{\vec{g}}_M:=\vec{f}^*M\vec{g}, \quad \textrm{with} \ \ 
   M:=\left(\begin{array}{ccc}
       h & 0 & 0 \\
       0 & h & 0 \\
       0 & 0 & g
   \end{array}\right).\label{matrix_M}
\end{equation}

The matrix $L$ in \eqref{L_operator} is known as the {dispersion matrix}, since its eigenvalues give the dispersion relation. Taking the definition in \eqref{L_operator} and replacing $\partial_\vec{x}$ by $\i\vec{k}$, we find that the solutions to the eigenvalue equation $L(\i\vec{k})\vec{b}(\vec{k})=\i\omega(\vec{k})\vec{b}(\vec{k})$ are given by
\begin{equation}
    \omega_0=0, \;\; \omega_{\pm}(\vec{k})=\pm\sqrt{f^2+gh|\vec{k}|^2},
\end{equation}
which is the usual dispersion relation for the rotating shallow-water model. The three eigenvectors are given by
\begin{equation}
    \vec{b}_0=\frac{1}{g\omega_\pm^2}\left(\begin{array}{c}
        -\i gl \\
         \i gk\\
         f
    \end{array}\right),\;\;\;\vec{b}_\pm=\frac{1}{\sqrt{2h}\abs{\omega_\pm} |\vec{k}|}\left(\begin{array}{c}
        \omega_\pm k+\i fl  \\
        \omega_\pm l-\i fk\\
         h|\vec{k}|^2
    \end{array}\right).\label{eigenvectors_appendix}
\end{equation}
These are orthonormal in the sense that
\begin{equation}
    \inner{\vec{b}_i}{\vec{b}_j}_M=\delta_{ij},\;\;\;i,j=0,\pm.\label{orthonormality}
\end{equation}
The zero eigenvalue $\omega_0$ corresponds to the vortical mode of the system. Since this mode is accounted for by the quasigeostrophic background flow, it does not appear in any of the following
derivation.
In addition to the right eigenvectors $\vec{b}_i$, we also use the left eigenvectors $\vec{c}_i$ which satisfy  
\begin{equation}
  \vec{c}_jL=\i\omega_j\vec{c}_j,  \quad \vec{c}_j=\vec{b}_j^*M\quad\textrm{and}\quad\vec{c}_i\vec{b}_j=\delta_{ij}.\label{left_eigenvectors}
\end{equation}

\subsection{Scaled Wigner transform}

It is convenient to rescale the space and time coordinates as $(\vec{x},t)\mapsto(\vec{x}/\varepsilon,t/\varepsilon)$, and define $\vec{\phi}_\varepsilon(\vec{x},t):=\vec{\phi}(\vec{x}/\varepsilon,t/\varepsilon)$. Under this rescaling, the equations of motion \eqref{dynamics} take the form
\begin{equation}
    \varepsilon\partial_t\vec{\phi}_\varepsilon+\big[L(\varepsilon\partial_\vec{x})+\sqrt{\varepsilon}N({\vec{x}}/{\varepsilon},\varepsilon\partial_\vec{x},{t}/{\sqrt{\varepsilon}})\big]\vec{\phi}_\varepsilon=0.\label{scaled-dynamics}
\end{equation}

In the new coordinates, the wavevector is $O(\varepsilon^{-1})$. To account for this and to obtain a well-defined Wigner transform in the limit $\varepsilon \to 0$, we use the rescaled definition
\begin{equation}
    W^\varepsilon(\vec{x},\vec{k},t)=\varepsilon^{-2}W(\vec{x},\vec{k}/\varepsilon,t)=\frac{1}{(2\pi)^2}\int_{\mathbb{R}^2}\exp^{\i \vec{k}\cdot\vec{y}}\vec{\phi}_\varepsilon(\vec{x}-\varepsilon{\vec{y}}/{2},t)\vec{\phi}_\varepsilon^*(\vec{x}+\varepsilon{\vec{y}}/{2},t)\d\vec{y}.\label{scaled-wigner}
\end{equation}
 
We note that this function remains positive semi-definite as $\varepsilon$ goes to zero, which is important as it is closely related to the energy of the system. Indeed it can readily be shown that 
\begin{equation}
   \int_{\mathbb{R}^2}W^\varepsilon(\vec{x},\vec{k},t)\d \vec{k}=\vec{\phi}_\varepsilon(\vec{x},t)\vec{\phi}_\varepsilon^*(\vec{x},t),
\end{equation}
and so the energy density is found from the Wigner transform of the wave fields as
\begin{align}
   \mathcal{E}(\vec{x},t)=\frac{1}{2}\int_{\mathbb{R}^2}\text{tr}\big(MW^\varepsilon(\vec{x},\vec{k},t)\big)\d\vec{k}.%&=\frac{1}{2}M_{ji}\int_{\mathbb{R}^2}W^\varepsilon_{ij}(\vec{x},\vec{k},t)\d\vec{k}
   \label{energy_density}
\end{align}

A useful alternative to \eqref{scaled-wigner} expresses the Wigner transform in terms of Fourier transforms as
\begin{equation}
    W^\varepsilon(\vec{x},\vec{k},t)=\varepsilon^{-2}\int_{\mathbb{R}^2}\exp^{\i\vec{p}\cdot\vec{x}}\hat{\vec{\phi}}_\varepsilon(-\tfrac{\vec{k}}{\varepsilon}-\tfrac{\vec{p}}{2},t)\hat{\vec{\phi}}_\varepsilon(-\tfrac{\vec{k}}{\varepsilon}+\tfrac{\vec{p}}{2},t)\d\vec{p},\label{scaled-duality}
\end{equation}
where 
\begin{equation}
\hat{\vec{\phi}}_\varepsilon(\vec{k},t)=\frac{1}{(2\pi)^2}\int_{\mathbb{R}^2}\exp^{\i\vec{k}\cdot\vec{x}}\vec{\phi}_\varepsilon(\vec{x},t)\d\vec{x} \quad \textrm{and} \quad 
   \vec{\phi}_\varepsilon(\vec{x},t)=\int_{\mathbb{R}^2}\exp^{-\i \vec{k}\cdot\vec{x}}\hat{\vec{\phi}}_\varepsilon(\vec{k},t)\d\vec{k}.
\end{equation}

\subsection{Wigner-transform evolution equation}

We obtain an evolution equation for the Wigner transform by differentiating \eqref{scaled-wigner} with respect to $t$ and substituting \eqref{scaled-dynamics}. This gives
\begin{align}
    \varepsilon&\partial_t W^\varepsilon(\vec{x},\vec{k},t)\\
    &=\frac{\varepsilon}{(2\pi)^2}\int_{\mathbb{R}^2} \exp^{\i\vec{k}\cdot\vec{y}}\Big(\Big[{\partial_t\vec{\phi}_\varepsilon(t,\vec{x}-\tfrac{\varepsilon\vec{y}}{2})}{}\Big]\vec{\phi}_\varepsilon^*(t,\vec{x}+\tfrac{\varepsilon\vec{y}}{2})+\vec{\phi}_\varepsilon(t,\vec{x}-\tfrac{\varepsilon\vec{y}}{2})\Big[{\partial_t\vec{\phi}_\varepsilon^*(t,\vec{x}+\tfrac{\varepsilon\vec{y}}{2})}{}\Big]\Big)\d{} \vec{y}\\
    &=\frac{-1}{(2\pi)^2}\int_{\mathbb{R}^2} \exp^{\i\vec{k}\cdot\vec{y}}\Big[ L (\varepsilon\partial_\vec{x})+\sqrt{\varepsilon}N(\tfrac{\vec{x}}{\varepsilon}-\tfrac{\vec{y}}{2},\varepsilon\partial_\vec{x})\Big]\vec{\phi}_\varepsilon(t,\vec{x}-\tfrac{\varepsilon\vec{y}}{2})\vec{\phi}_\varepsilon^*(t,\vec{x}+\tfrac{\varepsilon\vec{y}}{2})\d{} \vec{y}+\text{c.c.},
\end{align}
where c.c. denotes the complex conjugate of the term preceding it.  Rewriting $\vec{\phi}_\varepsilon$ and $N$ in terms of their Fourier transforms and making use of \eqref{scaled-duality}, we find that 
\begin{multline}
   \varepsilon\partial_tW^\varepsilon(\vec{x},\vec{k},t)+\overbrace{\Big[L(\i\vec{k}+\tfrac{\varepsilon}{2}\partial_\vec{x})W^\varepsilon(\vec{x},\vec{k})+\text{c.c.}\Big]}^{\textstyle :=\mathcal{Q}^\varepsilon W^\varepsilon}\\
   +\sqrt{\varepsilon}\Big[\underbrace{ \int\exp^{-\i\vec{p}\cdot{\boldsymbol\xi}}\hatt{N}(\vec{p},\i(\vec{k}+\tfrac{\vec{p}}{2})+\tfrac{\varepsilon}{2}\partial_\vec{x},\tfrac{t}{\sqrt{\varepsilon}})W^\varepsilon(\vec{x},\vec{k}+\tfrac{\vec{p}}{2})\d\vec{p} +\text{c.c.}}_{\textstyle :=\mathcal{P}^\varepsilon W^\varepsilon}\Big]=0.\label{wigner_evolution1}
\end{multline}
Scattering effects due to interactions of waves with the background flow are controlled by the third term, $\mathcal{P}^\varepsilon W^\varepsilon$. To derive it, we have introduced the fast-space variable $\boldsymbol\xi:=\vec{x}/\varepsilon$ and the Fourier transform $\hatt{N}$ of $N$ defined by
\begin{equation}
   N(\boldsymbol\xi,\cdot)=\int_{\mathbb{R}^2}\exp^{-\i \vec{p}\cdot\vec{\xi }}\hatt{N}(\vec{p},\cdot)\d\vec{p}. 
\end{equation}

We now derive the asymptotic limit of \eqref{wigner_evolution1} using a multiscale expansion. Defining the  intermediate time variable  $\tau:=t/\sqrt{\varepsilon}$ to cater for the time dependence of the streamfunction, we expand
\begin{equation}
    W^\varepsilon(\vec{x},\boldsymbol\xi,\vec{k},t,\tau)=W^{(0)}(\vec{x},\vec{k},t)+\sqrt{\varepsilon}W^{(1)}(\vec{x},\boldsymbol\xi,\vec{k},t,\tau)+\varepsilon W^{(2)}(\vec{x},\boldsymbol\xi,\vec{k},t,\tau)+{O}(\varepsilon^{3/2}),\label{wigner_multiscale}
\end{equation}
where we have anticipated that the leading-order term depends on the slow variables only. 
The differential operators are then expanded as
\begin{equation}
    \partial_\vec{x}\mapsto\partial_\vec{x}+\varepsilon^{-1}\partial_{\boldsymbol\xi},\;\;\;\partial_t\mapsto\partial_t+\varepsilon^{-1/2}\partial_\tau,
\end{equation}
where $\vec{x}$ and $\boldsymbol\xi$, $t$ and $\tau$ are  treated as independent variables, leading to the expansion
\begin{align}
    \mathcal{Q}^\varepsilon=\mathcal{Q}_0+\varepsilon\mathcal{Q}_1+O(\varepsilon^2),\;\;\;\mathcal{P}^\varepsilon=\mathcal{P}_0+\varepsilon\mathcal{P}_1+{O}(\varepsilon^2)
    \label{aa}
\end{align}
of the operators in \eqref{wigner_evolution1}.
It turns out that only the leading order term $\mathcal{P}_0$ is required for the derivation of the kinetic equation. 

The operators in \eqref{aa} can be written explicitly through their action on an arbitrary function $Z(\vec{x},\vec{\xi},\vec{k})$: 
\begin{align}
    \widetilde{\mathcal{Q}}_0Z(\vec{x},\vec{\xi},\vec{k})&= L(\i \vec{k}+\tfrac{1}{2}\partial_{\boldsymbol\xi})Z(\vec{x},\vec{\xi},\vec{k})+\text{c.c.} \label{Q_0} \\
    \widetilde{\mathcal{Q}}_1Z(\vec{x},\vec{\xi},\vec{k})&=\frac{1}{2\hspace{-0.2em}\i{}}\big[\nabla_\vec{k}L(\i \vec{k}+\tfrac{1}{2}\partial_{\boldsymbol\xi})\big] \cdot\nabla_\vec{x}Z(\vec{x},\vec{\xi},\vec{k})  +\text{c.c.}\label{Q_1}\\
    \widetilde{\mathcal{P}}_0Z(\vec{x},\vec{\xi},\vec{k})&=\int_{\mathbb{R}^2}\exp^{-\i \vec{p}\cdot\boldsymbol\xi}\hatt{N}(\vec{p},\i (\vec{k}+\tfrac{\vec{p}}{2})+\tfrac{1}{2}\partial_{\boldsymbol\xi},\tau)Z(\vec{x},\vec{\xi},\vec{k}+\tfrac{\vec{p}}{2})\d\vec{p}  +\text{c.c.}\label{P_0}
\end{align}
We have decorated the operators with a tilde to highlight the presence of $\partial_{\vec{\xi}}$ in their definition; the tildes will be removed whenever this dependence disappears.

Substituting the operators into \eqref{wigner_evolution1} gives us the evolution equation for the Wigner function as
\begin{equation}
    \Big[\frac{1}{\varepsilon}\widetilde{\mathcal{Q}}_0+\frac{1}{\sqrt{\varepsilon}}\Big(\widetilde{\mathcal{P}}_0+\frac{\partial}{\partial\tau}\Big)+\Big(\widetilde{\mathcal{Q}}_1+\frac{\partial}{\partial t}\Big)  \Big]W^\varepsilon(\vec{x},\boldsymbol\xi,\vec{k},t,\tau)=0.\label{Wigner_evolution}
\end{equation}
Introducing  the expansion \eqref{wigner_multiscale} then leads to a hierarchy of equations to be solved at each order in $\varepsilon$.

The leading-order equation is 
\begin{equation}
    \mathcal{Q}_0W^{(0)}=L(\i\vec{k})W^{(0)}(\vec{x},\vec{k},t)+\text{c.c.}=0.\label{Oe-1}
\end{equation}
The eigenvalues of $L$ are purely imaginary,  so this equation is satisfied by taking $W^{(0)}$ in the form of a linear combination of the eigenvectors of the dispersion matrix. Defining the matrices
\begin{equation}
    B_j(\vec{k})=\vec{b}_j(\vec{k})\vec{b}^*_j(\vec{k}),
\end{equation}
the leading order Wigner function is thus given by
\begin{equation}
    W^{(0)}(\vec{x},\vec{k},t)=\sum_{j=\pm}a_j(\vec{x},\vec{k},t)B_j(\vec{k}).\label{W0}
\end{equation}
The so-far undetermined amplitudes  $a_j(\vec{x},\vec{k},t)$ are real because the Wigner function is Hermitian.

At $O(\varepsilon^{-1/2})$, we find
\begin{equation}
    \widetilde{\mathcal{Q}}_0W^{(1)}(\vec{x},\boldsymbol\xi,\vec{k},t,\tau)=-\mathcal{P}_0 W^{(0)}(\vec{x},\vec{k},t),\label{Oe-1/2}
\end{equation}
where we have used that $\partial_\tau W^{(0)}=0$. 
To solve \eqref{Oe-1/2}, we rewrite $W^{(1)}$ in terms of its Fourier transform with respect to $\boldsymbol\xi$,
\begin{equation}
    W^{(1)}(\vec{x},\boldsymbol\xi,\vec{k},t,\tau)=\int_{\mathbb{R}^2}\exp^{-\i\vec{p}\cdot\boldsymbol\xi}\hatt{W}^{(1)}(\vec{x},\vec{p},\vec{k},t,\tau)\d\vec{p}.
\end{equation}
Substituting this into \eqref{Oe-1/2} yields
\begin{multline}
   L(\i(\vec{k}-\tfrac{\vec{p}}{2}))\hatt{W}^{(1)}(\vec{p},\vec{k})+\Big[L(\i(\vec{k}+\tfrac{\vec{p}}{2}))\hatt{W}^{(1)}(-\vec{p},\vec{k})\Big]^* + \theta\hatt{W}^{(1)}(\vec{p},\vec{k})\\
   =-\hatt{N}\big(\vec{p},\i(\vec{k}+\tfrac{\vec{p}}{2})\big)W^{(0)}(\vec{k}+\tfrac{\vec{p}}{2})-\Big[\hatt{N}\big(-\vec{p},\i(\vec{k}-\tfrac{\vec{p}}{2})\big)W^{(0)}(\vec{k}-\tfrac{\vec{p}}{2})\Big]^*,
\end{multline}
where we have suppressed dependencies on $\vec{x}$, $t$ and $\tau$ for conciseness. Following \cite{ryzhik}, we have introduced a regularisation parameter $\theta>0$  which will be taken to zero at a later stage.

Since the Wigner transform is Hermitian,  $\hatt{W}^{(1)}(\vec{p},\vec{k})=\hatt{W}^{(1)*}(-\vec{p},\vec{k})$. Using this, expanding $W^{(0)}$ according to \eqref{W0}, and pre- and post-multiplying the resulting expression by $\vec{c}_n(\vec{k}-\vec{p}/2)$ and $\vec{c}^*_m(\vec{k}+\vec{p}/2)$ (with $\vec{c}_n$
the left eigenvector defined in \eqref{left_eigenvectors}) gives
\begin{align}
   -\Big(\i(\omega_n(\vec{k}-&\tfrac{\vec{p}}{2})-\omega_m(\vec{k}+\tfrac{\vec{p}}{2}))+\theta\Big)\vec{c}_n(\vec{k}-\tfrac{\vec{p}}{2})\hatt{W}^{(1)}(\vec{p},\vec{k})\vec{c}_m^*(\vec{k}+\tfrac{\vec{p}}{2})\\
   =&\sum_{i=\pm}a_i(\vec{k}+\tfrac{\vec{p}}{2})\vec{c}_n(\vec{k}-\tfrac{\vec{p}}{2})\hatt{N}(\vec{p},\i(\vec{k}+\tfrac{\vec{p}}{2}))\vec{b}_i(\vec{k}+\tfrac{\vec{p}}{2})\vec{b}_i^*(\vec{k}+\tfrac{\vec{p}}{2})\vec{c}_m^*(\vec{k}+\tfrac{\vec{p}}{2})\\
   +&\sum_{j=\pm}a_j(\vec{k}-\tfrac{\vec{p}}{2})\vec{c}_n(\vec{k}-\tfrac{\vec{p}}{2})\vec{b}_j(\vec{k}-\tfrac{\vec{p}}{2})\vec{b}_j^*(\vec{k}-\tfrac{\vec{p}}{2})\hatt{N}^*(-\vec{p},\i(\vec{k}-\tfrac{\vec{p}}{2}))\vec{c}_m^*(\vec{k}+\tfrac{\vec{p}}{2}).
\end{align}
It is convenient to extract the linear dependence of $\hatt{N}$ on the streamfunction by defining a matrix $\hatt{U}(\vec{p},\i\vec{q})$ such that
\begin{equation}
    \hatt{N}(\vec{p},\i\vec{q},\tau)=\hatt{U}(\vec{p},\i\vec{q})\hatt{\psi}(\vec{p},\tau).\label{V=Upsi}
\end{equation}
 We now decompose $\hatt{W}^{(1)}$ using the vectors $\vec{b}_i(\vec{k})$, which form a complete basis, as
\begin{equation}
    \hatt{W}^{(1)}(\vec{x},\vec{p},\vec{k},t,\tau)=\sum_{m,n=\pm}\alpha_{mn}(\vec{x},\vec{p},\vec{k},t,\tau)\vec{b}_n(\vec{k}-\tfrac{\vec{p}}{2})\vec{b}_m^*(\vec{k}+\tfrac{\vec{p}}{2}).
\end{equation}
Using this along with the orthonormality of the eigenvectors we finally write the solution
\begin{multline}
    \hatt{W}^{(1)}(\vec{x},\vec{p},\vec{k},t,\tau)=\sum_{m,n=\pm}\Big[a_m(\vec{x},\vec{k}+\tfrac{\vec{p}}{2},t)\vec{c}_n(\vec{k}-\tfrac{\vec{p}}{2})\hatt{U}(\vec{p},\i(\vec{k}+\tfrac{\vec{p}}{2}))\vec{b}_m(\vec{k}+\tfrac{\vec{p}}{2})\\
    +a_n(\vec{x},\vec{k}-\tfrac{\vec{p}}{2},t)\vec{b}_n^*(\vec{k}-\tfrac{\vec{p}}{2})\hatt{U}^*(-\vec{p},\i(\vec{k}-\tfrac{\vec{p}}{2}))\vec{c}^*_m(\vec{k}+\tfrac{\vec{p}}{2})\Big]\frac{\vec{b}_n(\vec{k}-\tfrac{\vec{p}}{2})\vec{b}_m^*(\vec{k}+\tfrac{\vec{p}}{2})\hatt{\psi}(\vec{p},\tau)}{\i\big(\omega_m(\vec{k}+\tfrac{\vec{p}}{2})-\omega_n(\vec{k}-\tfrac{\vec{p}}{2})\big)-\theta},
\end{multline}
where we have taken into account that $\hatt{\psi }(\vec{p})=\hatt{\psi }^*(-\vec{p})$.
We note that this solution shows $W^{(1)}$ is linear in the random field $\psi$.

The slow evolution of the leading-order Wigner function $W^{(0)}$ is controlled by the $O(1)$ term in the expansion of  \eqref{Wigner_evolution}, given by
\begin{equation}
    -\widetilde{\mathcal{Q}}_0W^{(2)}=(\widetilde{\mathcal{P}}_0+\partial_\tau)W^{(1)}+(\mathcal{Q}_1+\partial_t)W^{(0)}.\label{Oe0}
\end{equation}
We assume that the random streamfunction is a stationary process in $\tau$ and homogeneous in $\vec{\xi}$, with zero mean, $\avg{\psi(\boldsymbol\xi,\tau)}=0$, and covariance 
\begin{equation}
    \avg{\psi(\boldsymbol\xi,\tau)\psi(\boldsymbol\xi',\tau)}=R(\boldsymbol\xi-\boldsymbol\xi'),
\end{equation}
where $\avg{\cdot}$ denotes an ensemble average, or equivalently an average over $\boldsymbol\xi$. In terms of Fourier transforms, this implies that
\begin{equation}
    \avg{\hatt{\psi}(\vec{p})\hatt{\psi}(\vec{p}')}=\hatt{R}(\vec{p})\delta(\vec{p}+\vec{p}')\label{power_spectrum},
\end{equation}
where the streamfunction power spectrum $\hatt{R}$ is the Fourier transform of $R$. The more familiar kinetic energy spectrum is then
\begin{equation}
\hatt{E}(\vec{k})=\abs{\vec{k}}^2\hatt{R}(\vec{k}).\label{energy_spectrum}
\end{equation}

We now take the average of \eqref{Oe0}.
 The slow time derivative term on the right-hand side disappears since $\avg{W^{(1)}}=0$. Since $\avg{\partial_\xi W^{(2)}}=0$,  $\avg{\widetilde{\mathcal{Q}}_0 W^{(2)}}=\mathcal{Q}_0 \avg{W^{(2)}}$, where the removal of the tilde corresponds to setting $\partial_{\vec\xi}$ to $0$ in $\mathcal{Q}_0$. This leads to 
\begin{equation}
    -{\mathcal{Q}}_0W^{(2)}=\avg*{\widetilde{\mathcal{P}}_0W^{(1)}+(\mathcal{Q}_1+\partial_t)W^{(0)}},\label{Oe0_averaged}
\end{equation}
an inhomogeneous version of \eqref{Oe-1}. 

The matrix $\mathcal{Q}_0$ has a non-trivial null space, spanned by the matrices $B_j(\vec{k})$;  the right-hand side of \eqref{Oe0_averaged} must therefore satisfy a solvability condition. Since  $\i\mathcal{Q}_0$ is self-adjoint with respect to the matrix inner product 
\begin{equation}
    \minner{X}{Y}:=\text{tr}(MX^*MY),
\end{equation}
this condition is obtained by applying $\minner{B_j(\vec{k})}{\cdot}$ to \eqref{Oe0_averaged}. We deal with the resulting terms one by one. First, by orthogonality and \eqref{W0} we have
\begin{equation}
    \minner{B_i}{\partial_tW^{(0)}}=\sum_{j=\pm}(\partial_ta_j)\minner{B_i}{B_j}=\partial_ta_i(\vec{x},\vec{k},t).
\end{equation}
Next,
\begin{align}
   \minner{B_i}{\mathcal{Q}_1W^{(0)}}&=\sum_{j=\pm}\frac{1}{2\hspace{-0.2em} \i{}}\minner{B_i}{(\nabla_\vec{k}L\cdot\nabla_\vec{x}a_j)B_j}+\text{c.c.}\\
   &=\sum_{j=\pm}\frac{1}{2\hspace{-0.2em}\i{}}\minner{B_i}{\nabla_\vec{k}(LB_j)-L\nabla_\vec{k}B_j}\cdot\nabla_\vec{x}a_j+\text{c.c.}\\
   &=\sum_{j=\pm}\frac{1}{2\hspace{-0.2em}\i{} }\minner{B_i}{\nabla_\vec{k}(\i\omega_j)B_j+(\i\omega_j-L)\nabla_\vec{k}B_j}\cdot\nabla_\vec{x}a_j+\text{c.c.}\\
   &=\nabla_\vec{k}\omega_i\cdot \nabla_\vec{x}a_i(\vec{x},\vec{k},t).
\end{align}

In order to evaluate the remaining term, we note that using \eqref{V=Upsi} and \eqref{power_spectrum}, we have
\begin{equation}
    \avg*{\hatt{N}_{\alpha\beta}(\vec{p},\i\vec{q})\hatt{N}_{\gamma\delta}(\vec{p}',\i\vec{q}')}=\hatt{U}_{\alpha\beta}(\vec{p},\i\vec{q})\hatt{U}_{\gamma\delta}(\vec{p}',\i\vec{q}')\hatt{R}(\vec{p})\delta(\vec{p}+\vec{p}'),\label{VV}
\end{equation}
where Greek indices are used for matrix elements to make the following derivation clearer, and  summation over repeated Greek indices is implied.

Expanding all terms, and making use of the delta function in \eqref{VV}, we have 
\begin{align}
  &  \minner{B_i}{\avg*{\widetilde{\mathcal{P}}_0W^{(1)}}}\\ &=-\iint\exp^{\i(\vec{p}+\vec{p}')\cdot\boldsymbol\xi}M_{\nu\rho}b_\rho^i(\vec{k})b_\sigma^{i*}(\vec{k})M_{\sigma\lambda}\avg*{\hatt{N}_{\lambda\mu}\big(\vec{p},\i(\vec{k}+\tfrac{\vec{p}-\vec{p}'}{2})\big)\hatt{W}^{(1)}_{\mu\nu}(\vec{p}',\vec{k}+\tfrac{\vec{p}}{2})}\d\vec{p}\d\vec{p}'+\text{c.c.}\\
   &=-\int\sum_{m,n=\pm} c^i_\lambda(\vec{k})\hatt{U}_{\lambda\mu}(\vec{p},\i(\vec{k}+\vec{p}))b_\mu^n(\vec{k}+\vec{p})\overbrace{c_\rho^m(\vec{k})b^i_\rho(\vec{k})}^{\delta^{im}}\hatt{R}(\vec{p})\\
   &\hspace{1em}\times  \frac{a_m(\vec{k})c_\alpha^n(\vec{k}+\vec{p})\hatt{U}_{\alpha\beta}(-{\vec{p}},\i\vec{k})b_\beta^m(\vec{k})+a_n(\vec{k}+\vec{p})\Big(c_\alpha^m(\vec{k})\hatt{U}_{\alpha\beta}({\vec{p}},\i(\vec{k}+\vec{p}))b_\beta^n(\vec{k}+\vec{p})\Big)^*}{\i\big(\omega_m(\vec{k})-\omega_n(\vec{k}+\vec{p})\big)-\theta} \d\vec{p}\\
   &\hspace{1em}+\text{c.c.}\\
   &=2\theta\Re\int\sum_{n=\pm}c_\lambda^i(\vec{k})\hatt{U}_{\lambda\mu}(\vec{\vec{k}'-\vec{k}},\i\vec{k}')b_\mu^n(\vec{k}')\hatt{R}(\vec{k}'-\vec{k})\\
   &\hspace{4em}\times\frac{a_i(\vec{k})c_\alpha^n(\vec{k}')\hatt{U}_{\alpha\beta}(\vec{k}-\vec{k}',\i\vec{k})b^i_\beta(\vec{k})+a_n(\vec{k}')\Big(c_\alpha^i(\vec{k})\hatt{U}_{\alpha\beta}(\vec{k}'-\vec{k},\i\vec{k}')b^n_\beta(\vec{k}')\Big)^*}{\big(\omega_i(\vec{k})-\omega_n(\vec{k}')\big)^2+\theta^2}\d\vec{k}',
\end{align}
where we have let $\vec{k}':=\vec{k}+\vec{p}$. Setting the regularisation parameter $\theta\to0$, we have that $\theta/(x^2+\theta^2)\to\pi\delta(x)$. This leads to a factor $\delta(\omega_i(\vec{k})-\omega_n(\vec{k}'))$ which indicates that scattering is restricted within a single branch of the dispersion relation, and so we may drop the sum over $n$. 

In order to evaluate this expression, we define
\begin{equation}
    c_\lambda(\vec{k})\hatt{U}_{\lambda\mu}(\vec{k}'-\vec{k},\i\vec{k}')b_\mu(\vec{k}'):=\alpha(\vec{k},\vec{k}')+\i\beta(\vec{k},\vec{k}').
\end{equation}
We find that 
\begin{equation}
    \alpha(\vec{k},\vec{k}')=\frac{\vec{k}'\times\vec{k}}{\omega^2\abs{\vec{k}}^2}\Big[(f^2+\omega^2)\vec{k}\cdot\vec{k}'-f^2\abs{\vec{k}}^2\Big],
\end{equation}
and
\begin{equation}
    \beta(\vec{k},\vec{k}')=\frac{f\omega}{\omega^2\abs{\vec{k}}^2}\Big[\abs{\vec{k}\times\vec{k}'}^2+\vec{k}\cdot\vec{k}'(\abs{\vec{k}}^2-\vec{k}\cdot\vec{k}')\Big],
\end{equation}
and we see that $\alpha(\vec{k},\vec{k}')=-\alpha(\vec{k}',\vec{k})$ and $\beta(\vec{k},\vec{k}')=\beta(\vec{k}',\vec{k})$. Using these symmetries makes it straightforward to show that
\begin{multline}
   \Re\; \Big(c_\lambda(\vec{k})\hatt{U}_{\lambda\mu}(\vec{k}'-\vec{k},\i\vec{k}')b_\mu(\vec{k}')\Big)\Big(c_\alpha(\vec{k}')\hatt{U}_{\alpha\beta}(\vec{k}-\vec{k}',\i\vec{k})b_\beta(\vec{k})\Big)\\=-(\alpha^2(\vec{k},\vec{k}')+\beta^2(\vec{k},\vec{k}')),
\end{multline}
and
\begin{multline}
   \Re\; \Big(c_\lambda(\vec{k})\hatt{U}_{\lambda\mu}(\vec{k}'-\vec{k},\i\vec{k}')b_\mu(\vec{k}')\Big)\Big(c_\alpha(\vec{k})\hatt{U}_{\alpha\beta}(\vec{k}'-\vec{k},\i\vec{k}')b_\beta(\vec{k}')\Big)^*\\=\alpha^2(\vec{k},\vec{k}')+\beta^2(\vec{k},\vec{k}').
\end{multline}
We may thus define the differential scattering cross section as
\begin{equation}
    \sigma(\vec{k},\vec{k}'):=2\pi\abs*{c_\lambda(\vec{k})\hatt{U}_{\lambda\mu}(\vec{k}'-\vec{k},\i\vec{k})b_\mu(\vec{k}')}^2\hatt{R}(\vec{k}'-\vec{k})\delta\big(\omega(\vec{k})-\omega(\vec{k}')\big)\label{app_cross_section}
\end{equation}
such that the amplitudes satisfy the kinetic equation
\begin{equation}
    \partial_t a+\nabla_{\vec{k}}\omega\cdot\nabla_\vec{x}a=\int_{\mathbb{R}^2}\sigma(\vec{k},\vec{k}')a(\vec{k}')\d\vec{k}'-\Sigma(\vec{k})a(\vec{k}),\label{trans_eq}
\end{equation}
with 
\begin{equation}
    \Sigma(\vec{k})=\int_{\mathbb{R}^2}\sigma(\vec{k},\vec{k}')\d\vec{k}'.
\end{equation}
Expanding the terms in \eqref{app_cross_section}, replacing the power spectrum with the energy spectrum using \eqref{energy_spectrum}, and expressing the delta function instead in terms of wave vectors, we find that the differential scattering cross section for the shallow-water system \eqref{SWEs} is given by \eqref{cross_section}.
% \begin{multline}
%     \sigma(\vec{k},\vec{k}')=\frac{2\pi}{gh\abs{\omega^3\vec{k}^5}}\Big\{\abs{\vec{k}\times\vec{k}'}^2\big[(\omega^2+f^2)\vec{k}\cdot\vec{k}'-f^2\abs{\vec{k}}^2\big]^2\\
%     +f^2\omega^2\big[\abs{\vec{k}\times\vec{k}'}^2+\vec{k}\cdot\vec{k}'(\abs{\vec{k}}^2-\vec{k}\cdot\vec{k}')\big]^2\Big\}\frac{\hatt{E}(\vec{k}-\vec{k}')}{\abs{\vec{k}-\vec{k}'}^2}\delta(\abs{\vec{k}}-\abs{\vec{k}'}),
% \end{multline}
% %\begin{multline}
% %    \sigma(\vec{k},\vec{k}')=\frac{\pi}{2}\frac{1}{\abs{\omega^3\vec{k}^5}}\Big\{\abs{\vec{k}\times\vec{k}'}^2\big[gh\abs{\vec{k}}^4-(f^2+\omega^2)(\abs{\vec{k}}^2-2\vec{k}\cdot\vec{k}')\big]^2\\
% %    +4f^2\omega^2\big[\abs{\vec{k}\times\vec{k}'}^2+\vec{k}\cdot\vec{k}'(\abs{\vec{k}}^2-\vec{k}\cdot\vec{k}')\big]^2\Big\}\hatt{R}(\vec{k}-\vec{k}')\delta(\abs{\vec{k}}-\abs{\vec{k}'}),
% %\end{multline}
% with $|\vec{k}\times\vec{k}'|=|kl'-k'l|$. We note that this is real, positive, and symmetric with respect to the interchange $\vec{k}\leftrightarrow\vec{k}'$.

%The symmetry of the differential scattering cross section under the interchange $\vec{k}\leftrightarrow\vec{k}'$ ensures energy conservation to leading order.
The conservation of the leading-order energy is established by noting that the equivalent of \eqref{energy_density} for the scaled Wigner function is
\begin{align}
    \mathcal{E}(\vec{x},t)&=\frac{1}{2}\text{tr}\int_{\mathbb{R}^2}M\big[W^{(0)}+\varepsilon^{1/2}W^{(1)}+{O}(\varepsilon)\big]\d\vec{k}\\
    &=\frac{1}{2}\int_{\mathbb{R}^2}\sum_{j=\pm}a_j(\vec{x},\vec{k},t)\d\vec{k}+{O}(\varepsilon^{1/2})\\
    &=\mathcal{E}_0(\vec{x},t)+{O}(\varepsilon^{1/2}).
\end{align}
Integrating \eqref{trans_eq} with respect to $\vec{k}$ and noting that the right-hand side  vanishes because of the symmetry $\sigma(\vec{k},\vec{k}')=\sigma(\vec{k}',\vec{k})$, we find leading order-energy conservation
\begin{equation}
    \partial_t\mathcal{E}_0+\nabla_\vec{x}\cdot\mathcal{F}_0=0,\label{energy_cons}
\end{equation}
with the leading-order energy flux
\begin{equation}
    \mathcal{F}_0(\vec{x},t):=\int_{\mathbb{R}^2} [\nabla_\vec{k}\omega(\vec{k})]a(\vec{x},\vec{k},t)\d\vec{k}.
\end{equation}

\section{Projection of simulation data onto modes}\label{app:modes}

We describe how the energy density $a(\vec{x},\vec{k},t)$ can be estimated from local Fourier transforms of the wave fields. Using the Fourier representation of the Wigner function  given in \eqref{scaled-duality} (with $\varepsilon=1$),
% \begin{equation}
%     W(\vec{x},\vec{k},t)=\int_{\mathbb{R}^2}\exp^{\i\vec{p}\cdot\vec{x}}\hatt{\vec{\phi}}(-\vec{k}-\vec{p}/2,t)\hatt{\vec{\phi}}^*(-\vec{k}+\vec{p}/2,t)\d\vec{p}.
% \end{equation}
it is easily verified that the projection property
\begin{equation}
    \int_{\mathbb{R}^2} W(\vec{x},-\vec{k},t)\d\vec{x}=|\hatt{\vec{\phi}}(\vec{k})|^2
\end{equation}
holds.
In order to discriminate between the energy contributions from the different modes, we expand the fields in the eigenvectors basis according to
\begin{equation}
    \hatt{\vec{\phi}}(\vec{k},t)=\sum_{j=\pm}A^{(j)}(\vec{k},t)\vec{b}_j(\vec{k}),
\end{equation}
so that the energy for each wavenumber is given by
\begin{equation}
    \tilde{\mathcal{E}}(\vec{k},t)=\frac{1}{2}\inner*{\hatt{\vec{\phi}}}{\hatt{\vec{\phi}}}_M=\frac{1}{2}(|A^{(+)}(\vec{k},t)|^2+|A^{(-)}(\vec{k},t)|^2),
\end{equation}
% The leading order energy in terms of the Wigner function is
% \begin{equation}
%   \avg*{\int  W(\vec{x},\vec{k},t)\d\vec{x}}=\int \sum_j a_j(\vec{x},\vec{k},t)\vec{b}_j(\vec{k})\vec{b}_j^*(\vec{k})\d\vec{x},
% \end{equation}
with $\inner{\,\cdot\,}{\,\cdot\,}_M$ as defined in \eqref{matrix_M}. Orthonomality of the eigenvectors means that we can extract the modal energy contributions by projection to find
\begin{equation}
   |A^{(j)}(\vec{k},t)|^2=\abs{ \inner{\vec{b}_j(\vec{k})}{\hatt{\vec{\phi}}(\vec{k},t)}_M}^2.
\end{equation}

In terms of the energy density, the leading-order energy is given by
\begin{equation}
    \tilde{\mathcal{E}}_0(\vec{k},t)=\frac{1}{2}{\int_{\mathbb{R}^2}\sum_{j=\pm}a_j(\vec{x},-\vec{k},t)\d\vec{x}}.
\end{equation}
Thus, we may track the energy from the `+' mode by projecting the Fourier transform of the wave fields:
\begin{equation} 
\int_{\mathbb{R}^2} a_+(\vec{x},-\vec{k},t) \d\vec{x}=\avg*{\abs{ \inner{\vec{b}_+(\vec{k})}{\hatt{\vec{\phi}}(\vec{k},t)}_M}^2}=\avg*{\abs{{\omega}{}\hatt{\eta}/{|\vec{k}|}}^2},
\label{energy_projection}
\end{equation}
with $\avg{\cdot}$ denoting the ensemble average. This relates to the leading-order energy density of a single wave mode to the sea-surface height.
% \begin{equation}
%     = h\abs{\hatt{\vec{u}}(\vec{k})}^2+g\abs{\hatt{\eta}(\vec{k})}^2.
% \end{equation}

% Energy density in spectral space of fields with $\omega=\omega_+$ is given by 
% \begin{equation}
%     \mathcal{E}_+(\vec{k},t)=\frac{1}{2}\int\minner*{B_+}{W(\vec{x},-\vec{k},t)}\d\vec{x}=\frac{1}{2}(h\abs{\hatt{\vec{u}}_+}^2+g\abs{\hatt{\eta}_+}^2)
% \end{equation}
% something like this.

%REFERENCES
\medskip
 \bibliographystyle{jfm}
\bibliography{mybib}
\end{document}